\documentclass[smallextended]{svjour3}       % onecolumn (second format)
\smartqed  % flush right qed marks, e.g. at end of proof
\usepackage{graphicx}
\usepackage{xspace}
\newcommand{\eat}[1]{}
\newcommand{\N}{G}
\newcommand{\MN}{\mathcal{G}}
\newcommand{\ABDUCE}{\textsf{ABACUS}\xspace}
\newcommand{\MCDSOLVER}{\textsf{MD}\xspace}
\newcommand{\GL}{\textsf{GL}\xspace}
\usepackage{algorithm}
\usepackage{algorithmic}
\usepackage[table]{xcolor}
\begin{document}
\title{\ABDUCE: frequent pAttern mining-BAsed Community discovery in mUltidimensional networkS
}
% \title{ABACUS: Apriori-BAsed Community discovery in mUltidimensional networkS
% }

\author{Michele Berlingerio         \and
        Fabio Pinelli \and
        Francesco Calabrese
}

\institute{M. Berlingerio \at
              IBM Research, Dublin, Ireland \\
              \email{mberling [at) ie.ibm.com}    
           \and
F. Pinelli \at
              IBM Research, Dublin, Ireland \\
              \email{fabiopin [at) ie.ibm.com}   
\and
F. Calabrese \at
              IBM Research, Dublin, Ireland \\
              \email{fcalabre [at) ie.ibm.com}    
}

\date{Received: date / Accepted: date}

\maketitle

\begin{abstract}
Community Discovery in complex networks is the problem of detecting, for each
node of the network, its membership to one of more groups of nodes, the
\emph{communities}, that are densely connected, or highly interactive,
or, more in general, \emph{similar}, according to a similarity
function. So far, the problem has been widely studied in
monodimensional networks, i.e. networks where only one
connection between two entities may exist. However, real networks are often
multidimensional, i.e., multiple connections between
any two nodes may exist, either reflecting different kinds of
relationships, or representing different values of the same
type of tie. In this context, the problem of 
Community Discovery has to be redefined, taking into account
multidimensional structure of the graph. We define a new concept of community that groups together nodes sharing
memberships to the same monodimensional communities in the different single dimensions. As we show,
such communities are meaningful and able to group nodes even if they might not be connected in any of the
monodimensional networks. 
We devise ABACUS (frequent pAttern mining-BAsed Community discoverer in mUltidimensional networkS), an
algorithm that is able to extract multidimensional communities based
on the extraction of frequent closed itemsets from monodimensional community memberships. 
Experiments on two different real multidimensional networks
 confirm the meaningfulness of the
introduced concepts, and open the way for a new class of algorithms
for community discovery that do not  rely on the dense connections
among nodes. 
\keywords{Community discovery \and Multidimensional Networks \and
  Social Network Analysis}
\end{abstract}

\section{Introduction}\label{sec:introduction}
Inspired by real-world scenarios such as social networks, technology
networks, the Web, biological networks, and so on, in the last years,
wide, multidisciplinary, and extensive research has been devoted to
the extraction of non trivial knowledge from networks. Predicting
future links among the nodes or actors of a network (\cite{germj}), detecting and studying the
diffusion of information among them (\cite{gurumine,misael}), mining frequent patterns of
nodes' behaviors (\cite{benvenuto,gspan,humanmining-survey}), are only
a few examples of 
tasks in the field of Complex Network Analysis, that includes, among all, physicians,
mathematicians, computer scientists, sociologists, economists and
biologists. 
The data at the basis of this field of research is huge,
heterogeneous, and semantically rich, and this allows to identify
many properties and behaviors of the actors involved in a
network. One crucial task at the basis of Complex Network Analysis is
Community Discovery, i.e., the discovery of a group of nodes densely
connected, or highly related. Many techniques have been proposed to
identify communities in networks (\cite{cdsurvey,bfsurvey}), allowing the detection of hierarchical
connections, influential nodes in communities, or just groups of nodes
that share some properties or behaviors. In order to do so, the
connections among the nodes of a network were so far posed at the center of
investigation, since they play a key role in the study of the network
structure, evolution, and behavior. 

Nowadays, most of the work done in the literature is limited
to a very simplified perspective of such relations, focusing only on
whether two nodes are connected or not, and possibly assigning a
strength to this connection. In the real world, however, 
this is not always enough to model all the available information about
the interactions between actors, including their multiple preferences, 
their multifaceted behaviors, and their complex interactions. 
While multiple types of connections among actors could still be
represented into a monodimensional network, by collapsing all
connections to one type and potentially affecting a measure of tie
strength, a more 
sophisticated analysis of the network structure, which could maintain
information on the semantic differences in how actors are connected,
would help all the techniques to provide more meaningful communities.  

To this aim, in this paper we deal with \emph{multidimensional
networks}, i.e. networks in which multiple connections may exist
between a pair of nodes, reflecting various interactions (i.e.,
dimensions) between them. Multidimensionality in real networks may be expressed by either
different types of connections (two persons may be connected because
they are friends, colleagues, they play together in a team, and so
on), or different quantitative values of one specific relationship
(co-authorship between two authors may occur in several different
years, for example).

\begin{figure}[t!]
\centering
\begin{tabular}{cc}
\hspace{-4mm} \includegraphics[scale=.21]{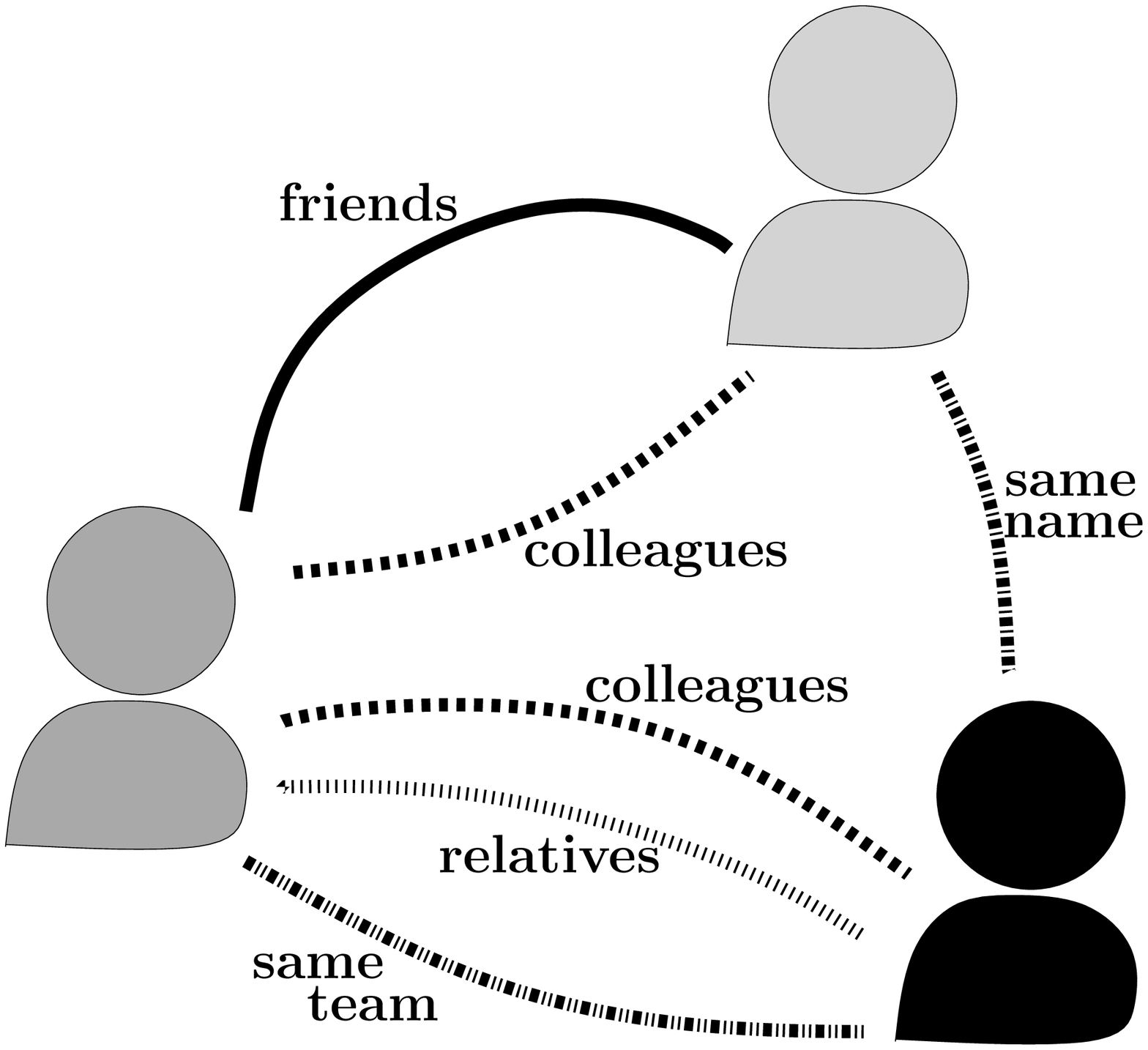} &\qquad
 \includegraphics[scale=.21]{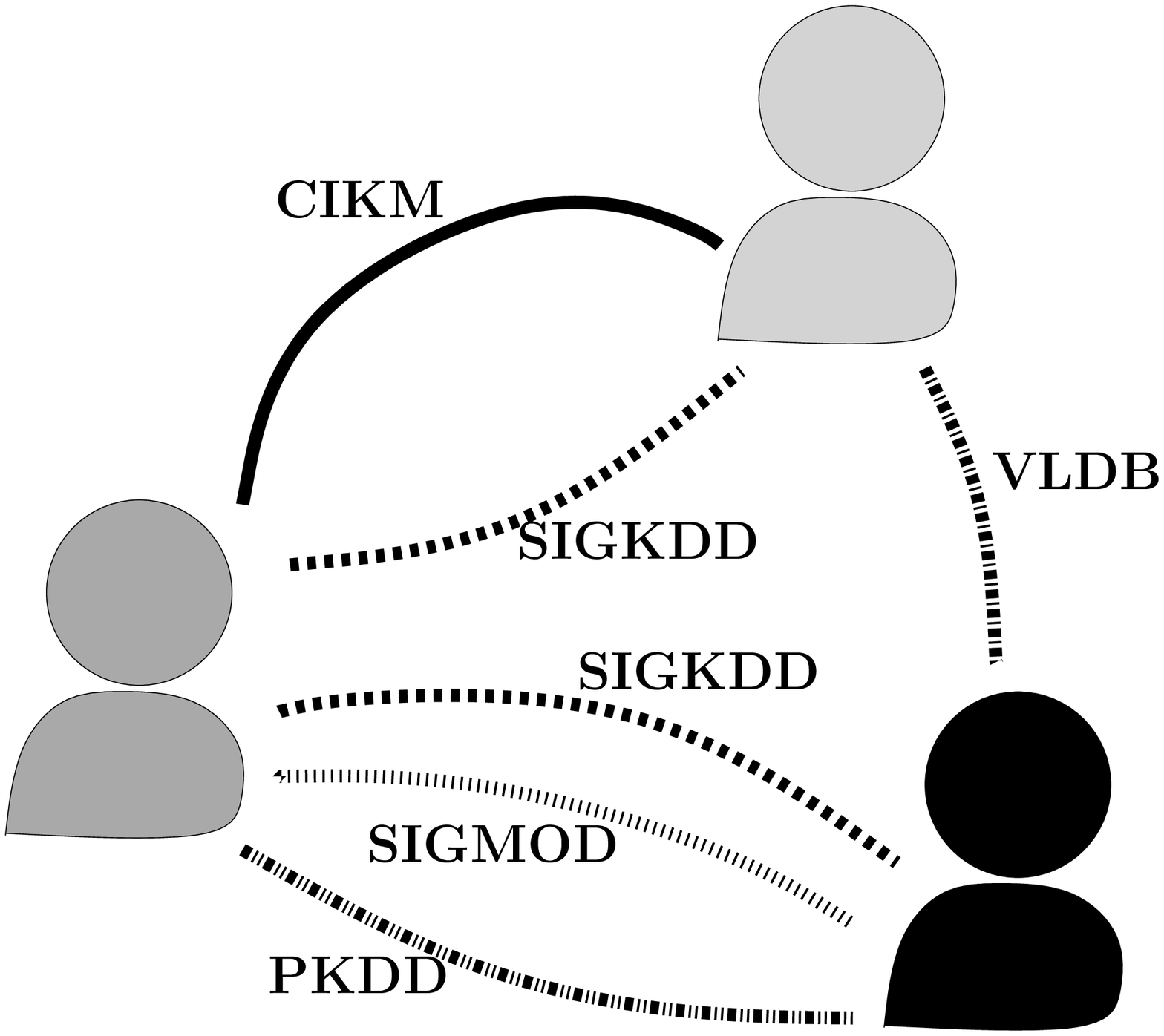} 
\end{tabular}
\caption{Example of multidimensional networks}
\label{fig:introexample}
\end{figure}

 This
 distinction is reported in Figure \ref{fig:introexample},
where on the left we have different types of links, while on the right
we have different values (conferences) for one relationship (for
example, co-authorship).
We can also distinguish between \emph{explicit}
or \emph{implicit} dimensions, the former being relationships
explicitly set by the nodes (friendship, for example), while the
latter being relationships inferred by the analyst, that may link two
nodes according to their similarity or other principles (two users may
be passively linked if they wrote a post on the same topic).
 
In this scenario, we deal with the problem of \emph{Multidimensional
  Community Discovery}, i.e. the problem of detecting communities of 
actors in a multidimensional network. We define a new concept of
multidimensional community that groups nodes sharing their membership
to the same monodimensional communities in the same single dimensions. This concept gives us
the possibility to leverage traditional monodimensional community
discovery algorithms.  
It then allows us to define the lattice of multidimensional
communities as function of the subset of dimensions for which the
monodimensional community memberships of nodes are shared.
Each multidimensional community can then be represented by the
associated subset of dimensions, providing a semantic meaning to the
community. Note that while the problem of finding cross-dimensional or cross-network
structures is not new
\cite{multicommunity,DBLP:conf/cikm/BerlingerioCG11,cliques06},
our definition of multidimensional community differs from the previous ones.
In fact, using this definition, a multidimensional community
could be unconnected, i.e. composed of nodes which are not directly connected in any of the dimensions.
This represents a complex phenomenon that can be seen in the real world: not all the people in
a social community are necessarily connected directly, and, if they
share their memberships in more than one dimension, they can be seen as
a (potentially unconnected) group of highly related (both positively
and negatively) people.

\begin{figure}[th!]
\centering
\includegraphics[scale=0.4,clip=true,trim=40 55 0 70]{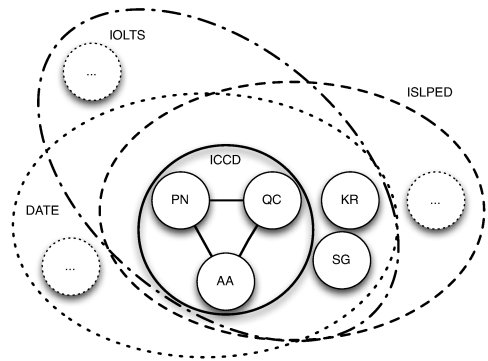} 
\caption{An example of a real multidimensional community found in DBLP by our
  algorithm ABACUS,
  that other methods are not able to detect. Nodes in the community:
  Amit Agarwal, Qikai Chen,
  Swaroop Ghosh,  Patrick Ndai,
  Kaushik Roy.}\label{fig:dblp_example_intro}
\end{figure}

We devise ABACUS (frequent pAttern mining-BAsed Community discoverer in
mUltidimensional networkS), an algorithm that extracts
multidimensional communities such as the one in Figure
\ref{fig:dblp_example_intro} working in four steps:
\begin{enumerate}
\item Each dimension is treated separately and monodimensional communities are extracted
\item Each node is labeled with a list of pairs (dimension, community the node belongs to in that dimension)
\item Each pair is treated as an item and a frequent closed itemset
  mining algorithm is applied
\item Frequent closed itemsets represent multidimensional communities described by the itemsets
\end{enumerate}
\ABDUCE is based on existing
monodimensional algorithms for community discovery (used as a
parameter), and on the extraction of frequent closed itemsets, that, in our scenario, represent the
multidimensional description of the communities.

Our main contribution can be then summarized as follows: we introduce
the new concept of multidimensional communities, and the \ABDUCE
algorithm to extract them
(Section \ref{sec:problem}); we show the applicability of \ABDUCE to
real world multidimensional networks (Section \ref{sec:analytics}),
together with a comparison with previous approaches to
the problem of community discovery in multidimensional network.

\section{Related work}\label{sec:related}
Detecting communities in networks has been studied from many angles.
Two comprehensive surveys on the topic can be found in \cite{bfsurvey,cdsurvey}.
From one side, a community has been defined as a set of nodes with a
 high density of links among them, and sparse connections
with nodes outside the community. The papers working with this quantitative definition
rely on information theory principles 
\cite{cct} or on the notion of modularity
\cite{clauset-modularity}, 
which is a function defined to detect the ratio between intra- and
inter-community number of edges. 
Modularity is widely
used in many works, and several algorithms have been proposed to
extract high modularity partitioning of a network: one of them is a 
greedy optimization able to scale up to networks with billions of
edges \cite{mod-unfolding}.
From another side, communities have been approached looking at the statistical properties of the
graph. In \cite{conga}, a framework for the detection of overlapping
communities, i.e. communities allowing the vertices to be in more than
one community, is presented. The framework is based on the ``split betweenness'' concept:
vertices and edges are ranked by their betweenness
centrality (the portion of shortest path in which they appear) and
then split in order to form a transformed network, where classical
algorithms can be used to detect communities. The resulting
communities are then merged in order to find overlaps.
Another class of approaches relies on the propagation in the network of
a label \cite{labelprop} or a particular definition of structure
(usually a clique \cite{kclique}). The first approach is known for
being a quasi linear solution for the problem, the
second one allows to find overlapping communities. 
One algorithm that maximizes quality and quantity measures on
its results is InfoMap \cite{infomap}, a random walk-based algorithm. 
An emerging novel problem definition can be found in
\cite{link-jaccard}, in which the authors state that community discovery
algorithms should not group nodes but edges, emphasizing the role of
the relation residing in a community.
Previously described methods have focused on both unweighted or
weighted graphs, but still considering the network as a
monodimensional entity. 
Only since recently, multidimensionality has started to be taken into account in network analysis.
A few examples of studies are: link 
prediction in networks with positive and negative links
\cite{lesko-multidimlinkpred} or in multidimensional networks
\cite{rossmultilinkpred}; a statistical analysis over
different kinds of relations in the same network in an online game
community \cite{szell-2010}; analysis of structural properties of
multidimensional networks \cite{foundations,wwwj} and its applications to
multidimensional hub analysis \cite{DBLP:journals/jocs/BerlingerioCGMP11}.

From a community discovery point of view, to the best of our
knowledge, the main approaches to take into account multiple dimensions are three. In \cite{onnela} the authors
extend the definition of modularity to fit the multidimensional
case, which they call ``multislice''. However, no definition of
``multidimensional community'' is provided, nor the approach
characterizes and analyzes the communities found. Instead, the authors
use the multidimensional information to extract monodimensional communities.
In \cite{edgecluster} the authors
create a machine learning procedure which detects the possible
different latent dimensions among the entities in the network and uses
them as features for the node classification algorithm. In other words,
they use the multidimensional labels of some nodes to infer labels for
other nodes, by means of edge clustering. Hence, 
multidimensionality is only present here in terms of node labels, but
the input network is not multidimensional according to our definition
(see Section \ref{sec:networks}), and the output is not in the form of
multidimensional communities.
In \cite{DBLP:conf/cikm/BerlingerioCG11}, a
possible formulation of community discovery and characterization in
multidimensional networks was given. A new measure was introduced to
capture the interplay among the dimensions, that makes 
multidimensional communities emerge even where the connections among
nodes reside in different dimensions.
In this paper, we approach the problem from a similar angle, but focus
on extracting communities using frequent itemset mining, and giving a semantic
description to each multidimensional community as the subset of
dimensions used to characterize it. Resulting multidimensional
communities may be different from the ones extracted in
\cite{DBLP:conf/cikm/BerlingerioCG11} and are navigable using the
lattice extracted in the frequent itemset mining process. 

Another work that deals with networks containing heterogenous
information, but not multiple dimensions, is presented in
\cite{SunYH09}, where the authors propose a method to generate
net-clusters using links across multi-typed objects. This approach
works on heterogeneous networks, i.e. networks where nodes may have
different types (e.g. papers or authors), and does not deal with
multidimensional networks, i.e. networks where edges may be of
different types and two nodes may be connected by multiple edges.

The authors of \cite{multicommunity} studied the problem
of community mining in multi-relational networks. The problem setting,
however, is different: the authors exploit the multi-relational links
to evaluate the importance of the relations based on labeled
examples, provided by a user as queries. Hence, they do not perform
community detection, but rather extract the importance of each
dimension for a given node, in the form of a weight.

The idea of applying closed frequent pattern mining to
multi-relational data is not new. In \cite{narymining}, the authors
extract all closed n-sets satisfying given piecewise (anti-)monotonic
constraints, from n-ary relations. In \cite{guns}, the authors
presented a framework for constraint-based
pattern mining in multi-relational databases, finding patterns
not under (anti-)monotonic and closedness constraints, expressed over
complex aggregates over multiple relations. 
However, both the works solve the technical
problem of finding the frequent closed patterns, but do not apply this
technique to the setting of multidimensional network analysis.

There are other works in the literature that deal with the extraction
of knowledge across networks. In \cite{cliques06} and
\cite{cliques06b}, for example, the authors deal with the problem of
finding cross-graph quasi-cliques. This problem can be seen as a
sub-problem of the one we deal with in this paper. However, our
concept of community is independent from the density of the 
connections among the nodes. Other two papers \cite{bouli,rigotti}
deal with the extraction of cliques with particular constraints: in
the first work, the authors search cliques that remain cliques over
time; in the second, cliques with homogeneous node attributes are
found. They however do not deal with the community discovery problem.

Based on all the above, we believe that two approaches may be considered
really related to our problem formulation, namely \cite{onnela} and
\cite{DBLP:conf/cikm/BerlingerioCG11}, thus we use these as baselines
for comparison in Section \ref{sec:analytics}.

\section{Multidimensional networks}\label{sec:networks}
In the world as we know it we can see a large number of interactions and
connections among information sources, events, people, or items,
giving birth to complex networks. Enumerating all the possible
networks detectable within our world, or their properties, would be
difficult due to their number and heterogeneity, and it is not the scope of this
paper. An excellent survey on complex networks can be found in 
\cite{newman}, where the author gives a good classification of
networks into \emph{social} (where, for example, we find on-line social network such
as Facebook), \emph{information} (such as for example citation
networks), \emph{technological} (among which we mention the power
grid, the train routes, or the Internet), and \emph{biological} (e.g.,
protein interaction networks) networks. 

While all the example networks presented in \cite{newman} are
monodimensional, in the real world it is possible to find many
multidimensional networks: transportation networks (transport means
are different dimensions), social networks (different online services
may be seen as different dimensions connecting the same users),
co-authorship networks (different venues as dimensions), constitute a
short, non-exhaustive list of possible real-world examples.

\subsection{A model for multidimensional networks}
In its classical definition, a network is defined as a structure that
is made up of a set of entities and connections among them. We want to
extend this definition by allowing connections of different kinds,
that we call \textit{dimensions}. 

We use a \emph{multigraph} to model a multidimensional network and its properties.
For the sake of simplicity, in our model we only consider undirected multigraphs
and since we do not consider node labels, hereafter we use  
\emph{edge-labeled undirected multigraphs}, denoted by a triple
$\mathcal{G}=(V,E,L)$ where: 
$V$ is a set of nodes; $L$ is a set of labels; $E$ is a set of
labeled edges, i.e. the set of triples $(u,v,d)$ 
where $u,v \in V$ are nodes and $d \in L$ is a label.
Also, we use the term \emph{dimension} to indicate \emph{label}, and
we say that a node \emph{belongs to} or \emph{appears in} a 
given dimension $d$ if there is at least one edge labeled with $d$
adjacent to it. We also say that an edge \emph{belongs to} or
\emph{appears in} a dimension $d$ if its label is $d$.
We assume that given a pair of nodes $u, v \in V$ and a label $d \in
L$ only one edge $(u,v,d)$ may exist. 
Thus, each pair of nodes in $\mathcal{G}$ can be connected by at most $|L|$
possible edges.  

\subsection{Real world dataset}
\begin{figure}[t!]
\centering\small
\begin{tabular}{cc}
\includegraphics[scale=0.42]{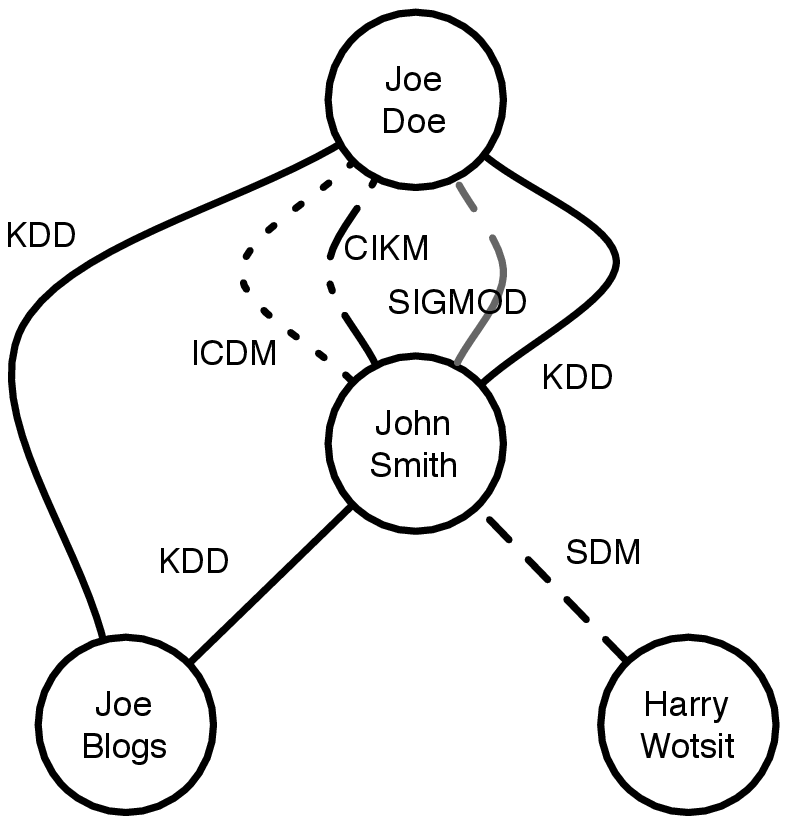}&
\includegraphics[scale=0.42]{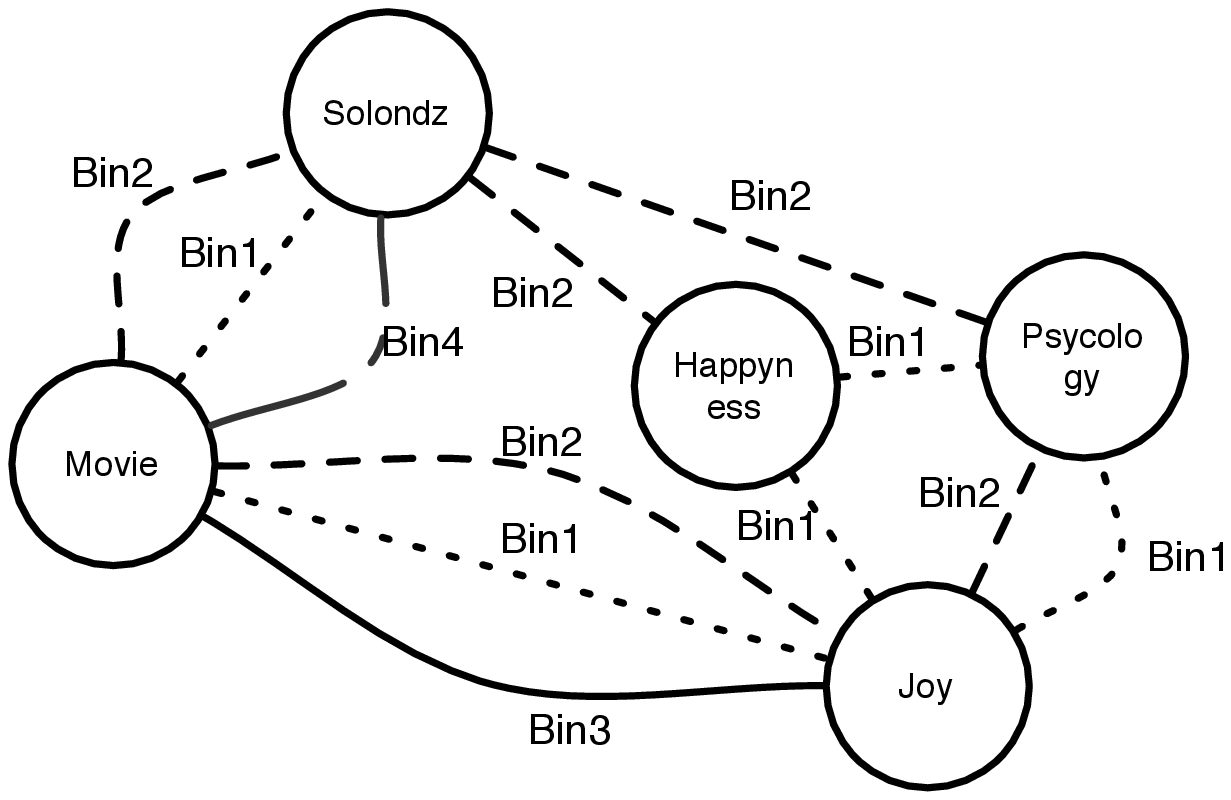}\\
(a) small DBLP extract & 
(b) small Query Log extract
\end{tabular}
 
\caption{Small extracts from the multidimensional DBLP and Query Log
  networks. Edge line styles correspond to dimensions in the network,
  i.e. distinct conferences for DBLP, and binned ranks for Query Log.}\label{fig:dblptoy}
\end{figure}
We created two multidimensional networks
from  the well known
  digital bibliography database 
  DBLP\footnote{http://dblp.uni-trier.de/xml} and from a search engine
  query log\footnote{http://www.gregsadetsky.com/aol-data}.

\begin{figure}[t!]
\centering
\begin{tabular}{cc}
\hspace{-4mm}
  \includegraphics[width=0.5\linewidth]{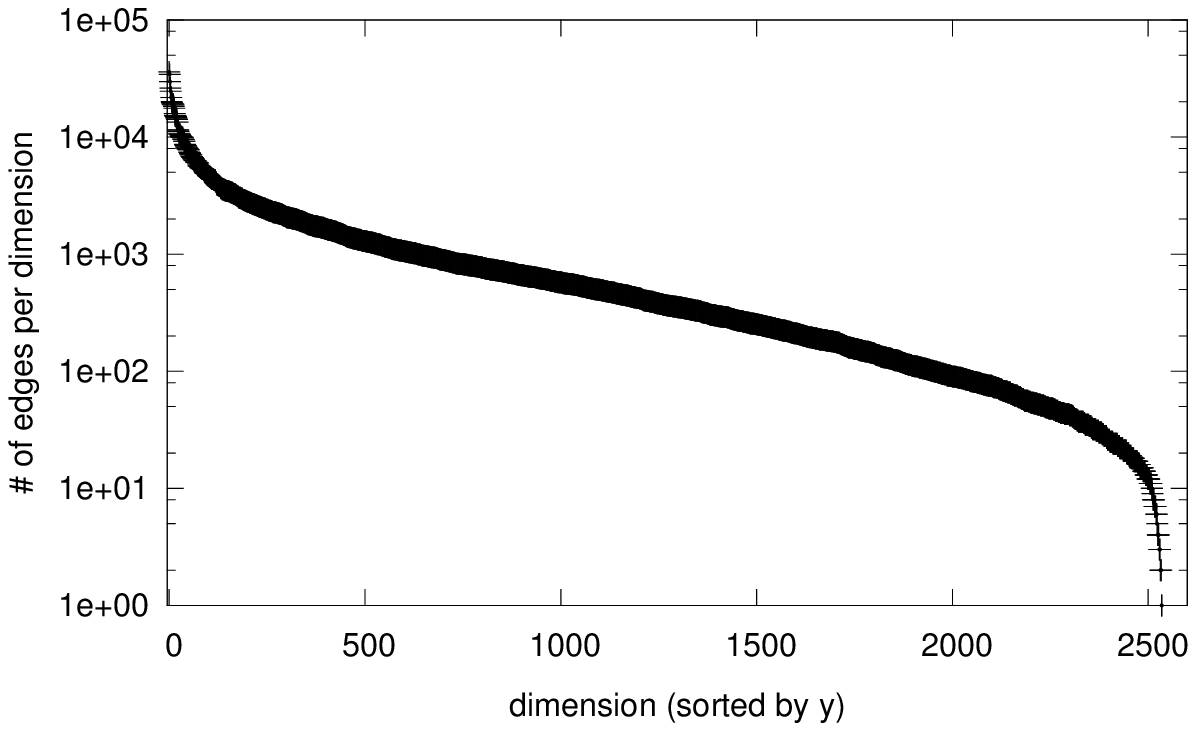}
& 
\includegraphics[width=0.5\linewidth]{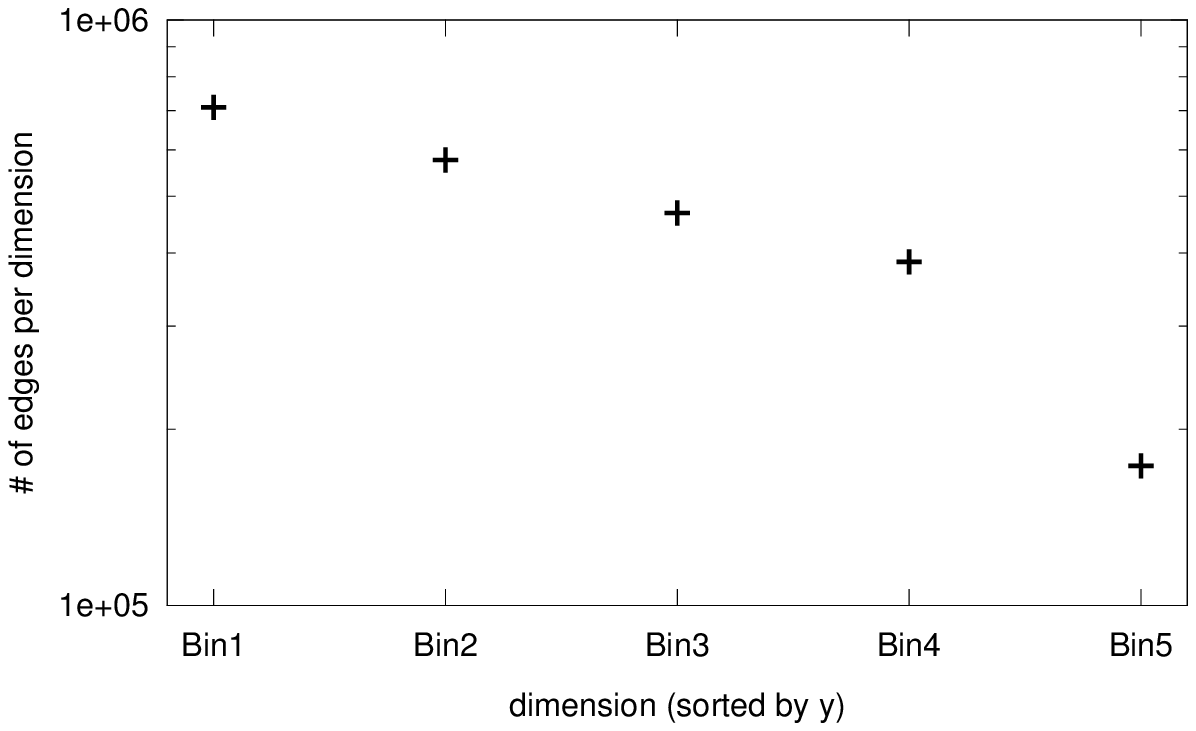}\\
(a) edges in DBLP dimensions & (b) edges in Query Log dimensions
\end{tabular}
\caption{
Distribution of the number of edges in each dimension in DBLP and
Query Log. Rank of the dimensions based on number of edges.}\label{fig:dblpedges}
\end{figure}

  \begin{itemize}
  \item \textbf{DBLP}
 We extracted author-author
  relationships if two authors collaborated in writing at least one
  paper. The dimensions of this network are defined as the venues
  in which the paper was published, resulting in 2,536 conferences that
  took place in years 2000-2010 (all the editions of a conference are
  considered as one dimension). As the network was created in year
  2012, we consider our temporal subset to be complete for years 2000-2010.
 We
weighted each edge by the number of papers published by the two
connected authors in the same conference (dimension).
The final network consisted of 558,800 nodes, connected by 2,668,497
edges in 2,536 dimensions. A small extract of this network is represented in Figure \ref{fig:dblptoy}(a).
Figure \ref{fig:dblpedges}(a) reports the
distribution of the number of edges per dimension (the dimensions are
sorted by the values of the y axis). High number of edges corresponds
to high number of editions of a conference and/or high number of
published papers and/or high co-authorship number per paper.

\item \textbf{QueryLog.} This network was constructed from a
query log of  
approximately 20 millions web-search queries submitted by 650,000
users, as described in \cite{aolquerylog}. 
 We extracted a word-word network of query terms (nodes),
connecting two words if they appeared together in a query. The dimensions are defined as the rank positions of the 
clicked results, grouped into six almost equi-populated bins: ``Bin1'' for rank 1, ``Bin2'' for ranks
2-3, ``Bin3'' for ranks 4-6, ``Bin4'' for ranks 7-10, ``Bin5'' for
ranks 11-500. Hence two words appeared together in a query for which 
the user clicked on a resulting url ranked \#4 produce a link in
dimension ``Bin3'' between the two words. We weighted each edge by the
number of queries in which the two connected words appeared together
in the same dimension.
The final network consisted of 131,268 nodes, connected by 2,313,224
edges in 5 dimensions. A small extract of this network is represented
in Figure \ref{fig:dblptoy}(b).
 Figure \ref{fig:dblpedges}(b) reports the
distribution of the number of edges per dimension. This network was
used in \cite{DBLP:journals/jocs/BerlingerioCGMP11} for tasks such as
query term disambiguation. More in general, word-word networks from
query logs have been used in the information retrieval literature to
mine the semantics of web search queries \cite{wordword,ricardo}
\end{itemize}

Note that we have weighted the edges of both DBLP and QueryLog to
preserve as much information as possible from the original data. These
edge weights may be taken into account by the monodimensional
community discovery algorithms to drive their search in a more
meaningful way, as they reflect the strength of the connections between nodes.

Following the classification in \cite{newman}, we took one social and
one information network, with different features (semantic: social vs
information network; number of dimensions: thousands vs five; number of
nodes: $\sim$600,000
vs $\sim$100,000, types of dimensions: categorical vs numerical
attributes). Although they do not cover the entire space of possible
networks, according to the classification in
\cite{newman}, our networks partially cover, with their structural
 characteristics and semantic, the spaces of social networks, information
networks, and technological networks.

\section{The \ABDUCE framework}\label{sec:problem}
In this section we present the core theoretical concepts of our
problem. After defining the types of communities we are seeking, we
 show how to map the problem of finding multidimensional communities
 to the problem of extracting frequent closed itemsets from community
 memberships, and then finally present \ABDUCE, the algorithm proposed
 to solve our definition of the problem.

\subsection{A new concept}
As said above, most of the existing approaches to the problem of
community discovery rely on a concept of community which is
structure-based. That is, nodes with dense connections (or high
interaction) are grouped together (in some cases, overlapping
communities are also discovered). In this paper, we change this
perspective. Let us start with a real-world example. In the WWW
context, nowadays it is very popular to be connected in services like
Facebook, Twitter, Google+ and, possibly, all of them. Each of these
services sees different communities that can be spotted within their sets
of nodes. As today many of the users have their online identities
replicated across the different social networks, it is very likely
that people sharing their membership in community $k$ in service $s$,
are also sharing their membership in community $k'$ in service
$s'$. Extending this, we can easily imagine that many communities
(especially small ones) would be exactly replicated across different
dimensions. 

In addition to this, there is another effect that can be detected in
the real world. Even within close circles of friends, it usually
happens to see pairs of people which are not directly connected. There can be
many reasons for this: they can be enemies, or potential friends not
yet connected, or there can be obstacles for their connection to setup
(in some example networks, spatial constraints may inhibit people living too far
away from connecting to each other). Yet, in these cases, two or more
persons can share their memberships to communities in different
contexts, or social networks, or, more in general, dimensions. 

Two nodes $A$ and $B$ can then end up being
logically connected by their shared memberships (say, to community $3$
in dimension Google+ and to community $4$ in dimension Facebook), but
never actually connected in any dimensions in which they
appear. This concept of logical connection here is crucial. While in
previous community discovery algorithms, monodimensional approaches
have a limited view of the rich set of connections residing within
nodes, disregarding the additional information provided by
multiple dimensions would be restrictive. Let us consider a
co-authorship graph in DBLP, where each conference is a different
dimension. Two persons in such network can be easily spotted to have
connections in conferences such as KDD, VLDB, and SDM, while they are
not connected, or not even present, in other dimensions such as AAAI,
or SIGGRAPH, and so on. This piece of information is usually lost in
traditional algorithms working on monodimensional networks, and,
unfortunately, weights do not help in conveying entirely this
additional knowledge. 

On the other hand, if we use the shared memberships as key concept for
connecting people (thus, not necessarily directly connected), we are
linking them logically, using the semantic residing in the dimensions.     

\subsection{From communities to itemsets}\label{sec:mapping}

Following the above idea, we can proceed as
follows. First, we can split a multidimensional network into
several monodimensional ones. We can then perform any existing technique for
monodimensional community discovery, obtaining, for each node of the
original network, a set of memberships to communities in each single 
dimension. We are now using the nodes as
\emph{transactions} of items, where an item is a pair $(dimension,
community)$ expressing the membership of the node in the various
dimensions. At this point, applying frequent pattern mining to find
\emph{frequent closed itemsets} \cite{closed} appears to be natural. There is, in
fact, a natural mapping of almost all the concepts in the frequent
closed itemset mining (FCIM) paradigm in our problem: nodes are transactions;
memberships are items; multidimensional communities are itemsets; the
support of an itemset is the number of nodes sharing that set of
memberships, and so on. Even the constraint-based paradigm
\cite{bonchivincoli} has a role in our problem: one can, in fact, use
constraints 
on the itemsets (eg. excluding/including specific items, computing any
monotonic or convertible measure on itemsets, and so on). For the sake
of simplicity, we reserve for future work this part of the problem,
and we focus only on the extraction of frequent closed itemsets.
In this new domain, it is also necessary to define concepts for a common understanding.
With the term \emph{support}, we intend the number of nodes that are members of a given 
multidimensional community. For instance, in the case of a co-authorship multidimensional network, 
two is the support of a multidimensional community formed by two
authors as members.
Moreover, the \emph{size} represents the number of different dimensions involved in a multidimensional community. 
Again in the multidimensional co-authorship network, two is the 
size of a multidimensional community composed of two dimensions such as two conferences.

\begin{figure*}[th!]
\centering
\begin{tabular}{c}
%\begin{minipage}{190pt} 
\includegraphics[clip=true,trim=20 0 0
  0,scale=0.4]{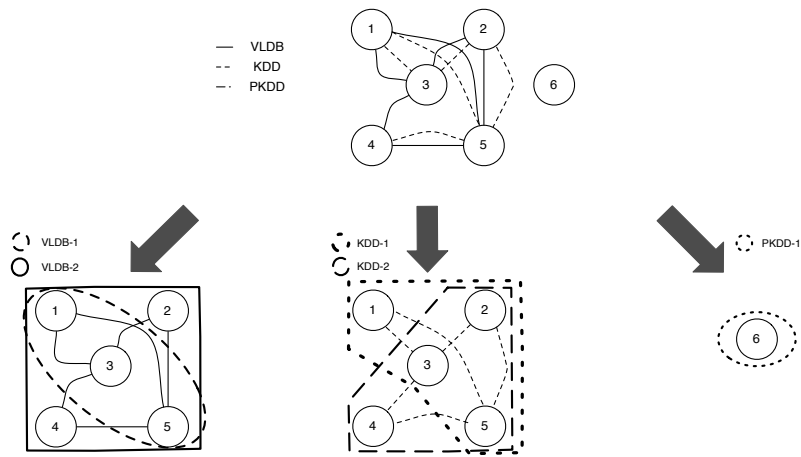} 
% \end{minipage} 
%&\hspace{1cm}
\end{tabular}
\caption{Run-through example: co-authorship network with two dimensions: KDD and VLDB (top), monodimensional overlapping communities (bottom)} 
\label{fig:runthrough1}
\end{figure*}

\begin{figure*}
\centering\tiny
\begin{tabular}{ccccc}
\hspace{-0.9cm}
\begin{minipage}{35pt}
%\begin{table}[hbc!]
\centering\small
\scalebox{0.6}{
\begin{tabular}{|c|c|}\hline
TID & ITEMS\\\hline
1 & ABC\\
2 & BCE\\
3 & ABCE\\
4 & BE\\
5 & ABCE\\
6 & D\\
\hline
\end{tabular}
}
 \end{minipage} 
& $\Rightarrow$
&\hspace{-0.5cm}
\begin{minipage}{209pt} \includegraphics[scale=0.38]{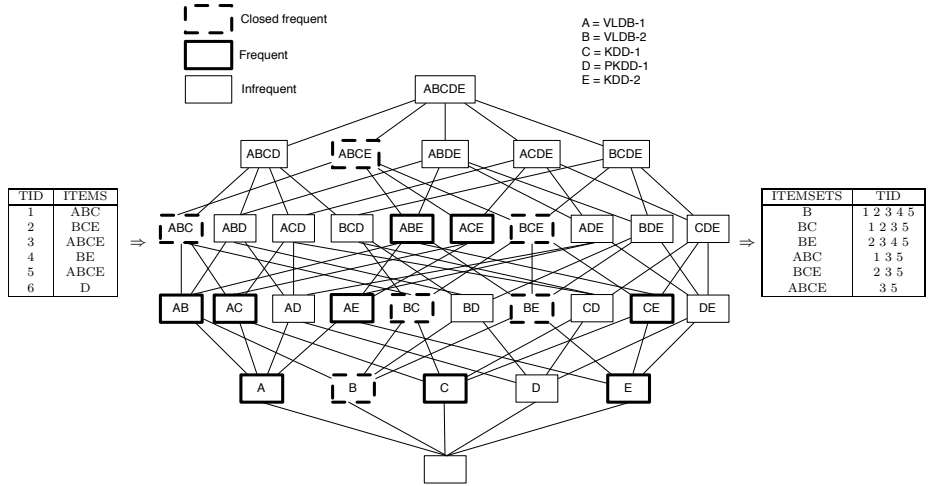} \end{minipage}
& $\Rightarrow$
&\hspace{-0.5cm}
\begin{minipage}{40pt}
%\begin{table}[hbc!]
\centering\small
\scalebox{0.6}{
\begin{tabular}{|c|c|}\hline
ITEMSETS & TID\\\hline
B    & 1 2 3 4 5 \\
BC   & 1 2 3 5\\
BE   & 2 3 4 5\\
ABC  & 1 3 5\\ 
BCE  & 2 3 5\\
ABCE & 3 5\\
\hline
\end{tabular}
}
 \end{minipage} 

  \\

\end{tabular}
\caption{Run-through example: set of monodimensional communities
  (items) associated to each node (transaction), on the left; lattice of
 multidimensional communities extracted using an algorithm for mining
 frequent closed itemsets, in the middle; resulting frequent closed
 itemsets (multidimensional communities) and supporting transactions
 (nodes) on the right.} 
\label{fig:runthrough}
\end{figure*}

Let us
follow a run-through example of our search strategy. Figure
\ref{fig:runthrough1} describes our toy input network
(top), consisting of six nodes, connected in three different
dimensions (KDD, VLDB and PKDD). From the top image to the ones below, we perform
two steps: first, we split the multidimensional network into three
monodimensional ones; then, we perform the community discovery on each
of them. The algorithm finds two different communities (highlighted by
different line styles) in the VLDB and KDD dimensions, and
one formed by a single node in the PKDD dimension. Note
that, as this is just meant to provide an example to guide the reader
through the steps of our methodology, we did not run a real community
discovery algorithm here, and instead built the communities such that
the resulting output would contain all the features we want to explain
by means of this example. In particular, we imagined to be running an overlapping
monodimensional community discoverer, assigning communities also to
single nodes (node 6).
The output of this process is represented on the left of Figure \ref{fig:runthrough} 
that shows the list of transactions that it is possible to
build from the memberships of the five nodes. The central part of Figure
\ref{fig:runthrough} shows then how the lattice of
multidimensional communities is created. We see how the community
found in the PKDD dimension gets cut due to a minimum
support threshold $\sigma=2$. In bold black we have the frequent
itemsets, while with bold dashed line we highlighted the closed frequent
ones. Finally, we see that the closed frequent itemsets clearly
summarize the entire set of frequent itemsets found, so it would be
redundant to return also non-closed items. In the right part of the
figure we show the final output. The 
first community is found with a single membership to B,
i.e. VLDB-2. This is clearly a monodimensional community, that our
algorithm is still able to extract. The last community, formed by
nodes 3 and 5, shows how we are able to extract communities of nodes
that were not necessarily connected in the initial input. Indeed,
nodes 3 and 5 are unconnected in all the dimensions of the example. We
want to emphasize here that the ability to find such communities is
not given by the type of monodimensional community discoverer, but it
is due to the mapping to the frequent pattern mining paradigm, and the
fact that our concept of multidimensional communities groups together
nodes sharing memberships to the same monodimensional communities in
the same dimensions. Nodes 3 and 5, in fact, share their memberships
to the communities VLDB-1, VLDB-2, KDD-1 and KDD-2.  

\subsection{The \ABDUCE algorithm.}
% \subsection{\ABDUCE: a frequent pAttern mining-BAsed Community discoverer in mUltidimensional  networkS }

Algorithm \ref{alg:mcd} is the core of our approach. It takes as input
three parameters: the multidimensional network $\MN$, a
monodimensional algorithm for community discovery $CD$, and a minimum
support threshold $\sigma$. The algorithm works by building a set of
transactions $memberships$ that, for each node $n$, record a set of
pairs $(i,j)$ representing memberships of node $n$ to community $j$ in
dimension $i$. Note that if $CD$ is able to
find overlapping communities, one node may have more
than one pair associated to a specific dimension. This would result in
more possible combinations, i.e. more different items, thus an higher
number of resulting communities. However, this does not change the
type of communities that \ABDUCE may find, namely groups of nodes
sharing memberships to the same monodimensional communities in the
same dimensions. 
Thus, for the sake of simplicity, and without lack
of generality, in the rest of the paper we show experiments conducted
with a 
non-overlapping community discovery algorithm. 

Note also that $CD$ may or may not take into account edge weights to
drive the search for communities in a more meaningful way. As we have
weighted our networks presented in Section \ref{sec:networks}, in our
experimental evaluation in Section \ref{sec:analytics} we use an
algorithm that takes into account edge weights.

In line $4$ the
function $\phi$ is called to split the multidimensional network into a
set of monodimensional ones, by replicating each node into each of the
dimensions in which it has at least one edge, and adding to it all of
its adjacent edges in their corresponding dimensions. Each dimension
is then processed as a separate network $G_i$ by $CD$ in a for loop, returning a
different set of communities per dimension. In lines $6-8$, for each
node in each community, its memberships are updated with the pair
$(dimension, community)$, building a set of transactions (one per
node). The function $map$ returns a unique item code for its
argument. Such set is then passed to the frequent closed itemset miner
($FCIM$) in line $11$, together with
a threshold of minimum support, and the resulting set of frequent
closed 
itemsets % and their corresponding transaction ids 
are returned, constituting the multidimensional description of each
community. In Section \ref{sec:analytics} we show how, by using an
implementation of $FCIM$ returning also the transaction ids of each
itemset, we also get the 
% (as a set of pairs $(dimension, community)$)
% , the latter
% constituting the 
 set of nodes contained in each community (i.e., the
ids of the transactions supporting the frequent closed itemset).

\begin{algorithm}[t]\caption{\label{alg:mcd}\ABDUCE}
\small
\begin{algorithmic}[1]

\REQUIRE{$\MN,CD,\sigma$}
%\ENSURE{set of multidimensional communities $\mathcal{C}$ }
\smallskip
\FORALL{$n\in nodes(\MN)$}
\STATE $memberships[n]=\emptyset$
\ENDFOR
%\STATE $\{\N\}\leftarrow\phi(\MN)$
%\STATE $\mathcal{C}\leftarrow\phi'(C)$
%\FORALL{$\N_i\ \in\ \{\N\}$}
\FORALL{$\N_i\ \in\ \phi(\MN)$}
%\STATE $\{c\}\leftarrow CD(\N_i)$
\FORALL{$c_j \in\ CD(\N_i)$}
\FORALL{$n\in nodes(c_j)$}
\STATE $memberships[n]\leftarrow memberships[n]\cup map((i,j))$
\ENDFOR
\ENDFOR
\ENDFOR 
\STATE $ \mathcal I \leftarrow FCIM(memberships,\sigma)$
\RETURN $\mathcal I$

%\STATE $(\{i\_set\}, \{t\_id\})\leftarrow apriori(memberships,\sigma)$
%\RETURN $(\{i\_set\}, \{t\_id\})$
\end{algorithmic}
\end{algorithm}

The complexity of \ABDUCE is directly inherited by the complexity of
the algorithm used for $FCIM$, and by that of the method for
monodimensional community discovery. The additional complexity
introduced by \ABDUCE, in fact, resides only in the problem-mapping
phase, where we perform a linear scan of the list of communities found and we prepare the
input for $FCIM$. We then refer to the corresponding papers for
discussion on the complexity, although in Section \ref{sec:scalability}
we present an empirical evaluation of the complexity of \ABDUCE.

\section{Case study on DBLP and Query Log}\label{sec:analytics}
\subsection{Tools}
We have implemented \ABDUCE in c++, making use of the
igraph\footnote{http://igraph.sourceforge.net} library. 

As $CD$
parameter, we use the community discovery algorithm based on label
propagation \cite{labelprop}, that takes into account edge weights. This algorithm is well known to be
scalable, and, as a result, our running times to process the
network were considerably low (a few seconds up to the creation of the
transaction file, plus a few minutes to perform frequent closed
itemset mining, see Section
\ref{sec:comparison} for running times). In all the
experiments we set the minimum support threshold to $2$, in order to
capture all the possible connections among nodes. 

We chose an efficient implementation of 
Eclat \cite{eclatb} as frequent itemset miner, with options to return both frequent
closed itemsets and list of supporting transactions for every itemset.

Note that many other choices
are possible for the $CD$ and the frequent itemset mining steps and that, for the sake of
simplicity and presentation, we 
only report the results obtained by the above choice. % We leave for
% further research the investigation on the sensitivity of our approach
% to the choice of the implementation of these two steps.
Note also that
while the choice for the frequent closed itemset mining implementation is usually mainly
driven by scalability issues, selecting a different algorithm for
community discovery may lead to very different communities. The debate
on which algorithm to choose is however out of scope in this paper,
and we refer to Section \ref{sec:related} and to the surveys on community discovery for driving the
reader to the best choice for this step, which is mainly driven by the
final application \cite{bfsurvey,cdsurvey}. Moreover, despite
the possibility of returning different types of communities, we want
to emphasize that the ability to return potentially unconnected
communities is given by the mapping of the problem as described in
Section \ref{sec:problem}, and \emph{not} to the choice of $CD$. In
fact, as said above, our multidimensional communities represent nodes that share
memberships to the same monodimensional communities in the same
dimensions. This concept is not tied to the fact that the nodes must
be directly connected in all the dimensions found in the
multidimensional community.
 
All the experiments were performed on a laptop equipped with an Intel
i7 processor at 2.2GHz, with 4GB of RAM.  

\subsection{Experiments}
We performed our experiments following three questions related to our
problem:
\begin{itemize}
% \item[\textbf{Q1.}] Quantitative evaluation: can we spot regularities
%   and anomalies in the solution space of the frequent itemset miner? Can we
%   measure and identify such anomalies?
% \item[\textbf{Q1.}] Quantitative evaluation: can we spot regularities
%   and anomalies in the solution space of the apriori algorithm? Can we
%   measure and identify such anomalies?
\item[\textbf{Q1.}] Quantitative evaluation: given the high number of
  resulting communities, how can we easily reduce the patterns to select only a set of
  meaningful ones?
\item[\textbf{Q2.}] Are there relational
  dependencies between our concept of communities and structural
  properties of them?
\item[\textbf{Q3.}] Qualitative evaluation: among the communities
  found, are there any relevant ones? Can we reason on the
  multidimensional density of the connections within the communities? 
\end{itemize}

\begin{figure}[b!]
\centering\tiny
\begin{tabular}{c}
\includegraphics[width=0.72\linewidth,angle=0]{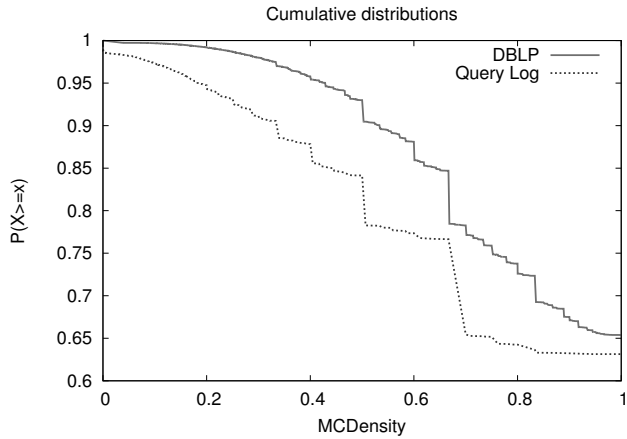}\\
\end{tabular}
\caption{Cumulative distributions of Multidimensional Community
  Density (MCD) for the two networks}\label{fig:supportFreq-dblp}
\end{figure}

In order to answer the above questions, we define a simple and easy to
compute measure
of connectedness within communities.
The Multidimensional Community Density (MCD) is then the number of
edges in a community normalized by the maximum possible for that
community, or, in formula:
\begin{equation}
\frac{\#edges}{ndim\times\frac{\#nodes\times(\#nodes -1)}{2}}
\end{equation}
where $ndim$ is the number of different dimensions found in the community.

% To answer \textbf{Q1}, we looked at the distributions of the
% communities. 
% By applying \ABDUCE to the DBLP multidimensional network,
% we obtained 484,833 multidimensional communities with at least size 2,
% while we found 14,415 communities in Query Log (the high unbalance is
% due to the high unbalance of the number of dimensions in the two
% datasets). We address later in the paper the problem of filtering this
% large number of resulting communities.

% Figure \ref{fig:supportFreq-dblp} shows the cumulative
% distributions of MCD on the resulting communities. As we can see, the line corresponding to Query
% Log is globally under the one corresponding to DBLP. One possible
% explanation for this is the higher number of edges per dimension in
% Query Log (see Figure \ref{fig:dblpedges}).

% The distributions of the support (MCS, number of nodes in a
% community) and the size of the patterns (see
% Figure \ref{fig:quattro}, top row for MCS, bottom row for the size)
% were in line with the literature of the applications of Frequent Pattern Mining. MCS ranged from 2 to 
% 216 for DBLP, and from 2 to 70,303 for Query Log. The size ranged from
% 1 to 63 for DBLP, and from 1 to 5 for Query Log (note however that we
% are not very interested in the results with size equal to 1, as they
% are truly monodimensional communities). 

\begin{figure}
\centering
\begin{tabular}{cc}
\hspace{-3mm}\includegraphics[width=0.5\linewidth,angle=0]{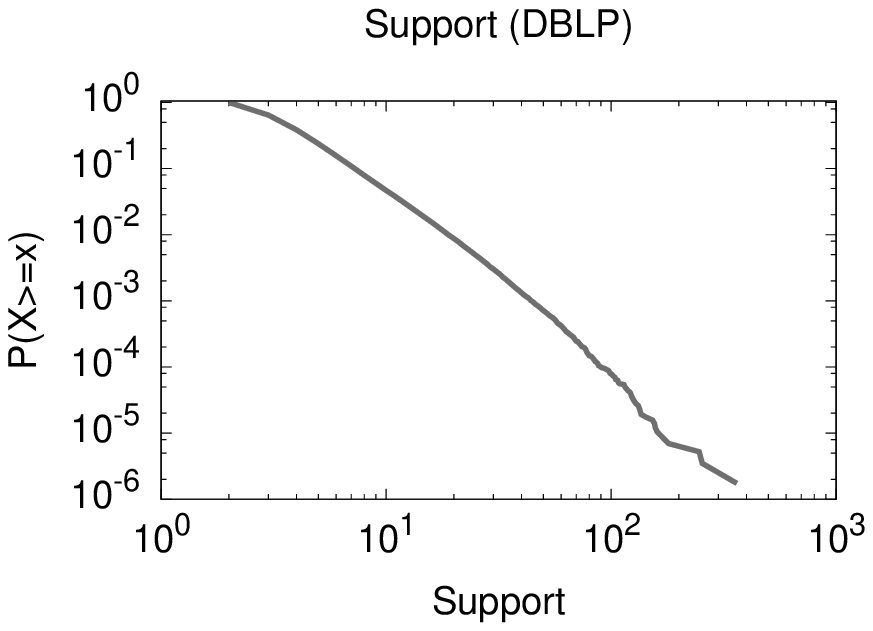}
&
\hspace{-2mm}\includegraphics[width=0.5\linewidth,angle=0]{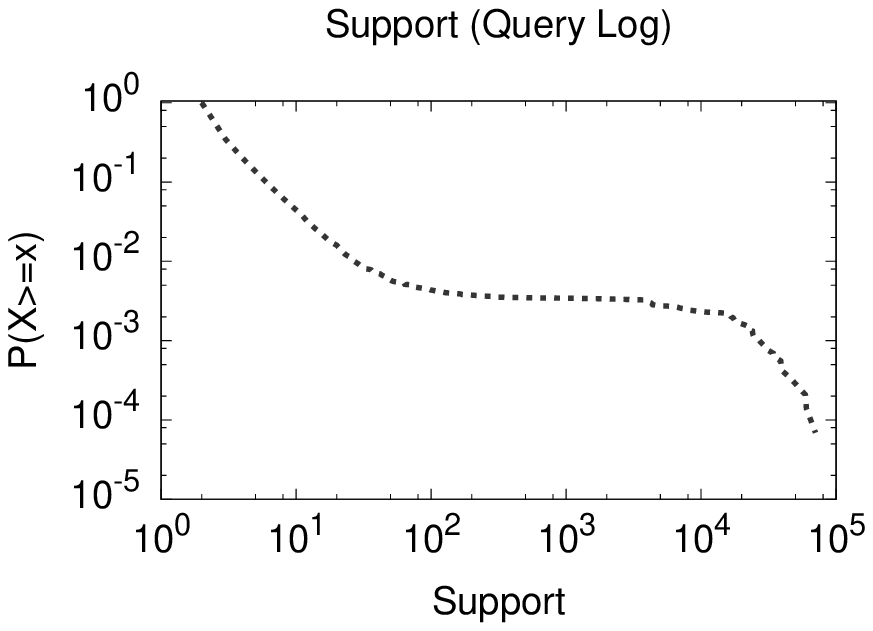}\\
\hspace{-2mm}(a) Distribution of the support in DBLP & 
\hspace{-2mm}(b) Distribution of the support in Query Log \\\ \\
\hspace{-3mm}\includegraphics[width=0.5\linewidth,angle=0]{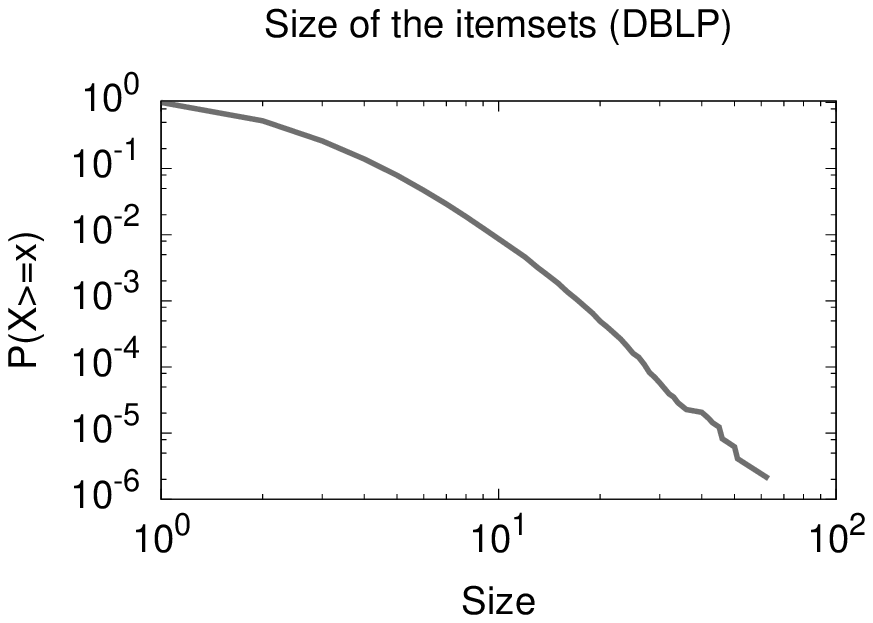}
&
\hspace{-2mm}\includegraphics[width=0.5\linewidth,angle=0]{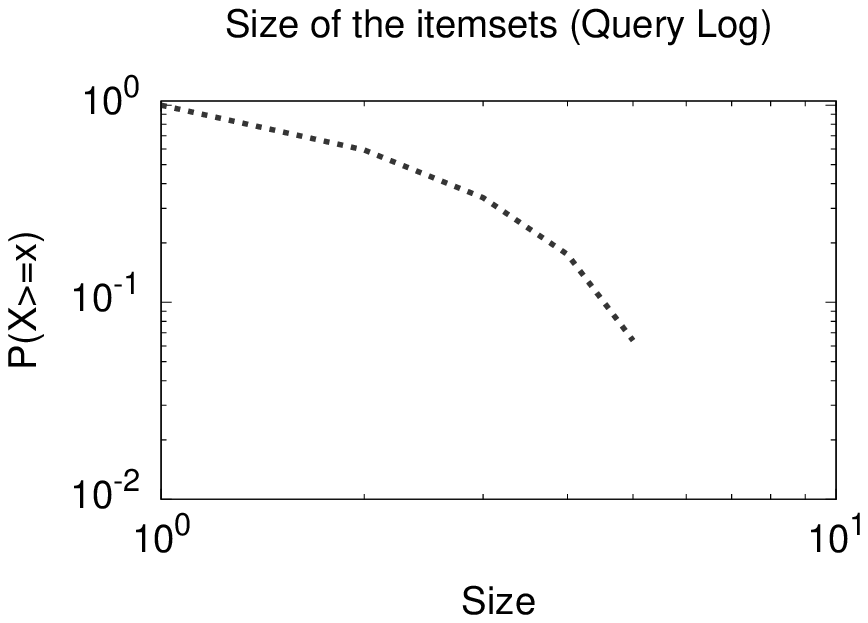}\\
\hspace{-2mm}(c) Distribution of the size in DBLP
& 
\hspace{-2mm}(d) Distribution of the size in Query Log
\end{tabular}
\caption{Cumulative distribution of support (top row) and size of the
  itemsets (bottom row), for DBLP (left column) and Query Log (right column)}\label{fig:quattro}
\end{figure}

Let us answer \textbf{Q1}. The frequent pattern mining
literature reports that the problem of finding few relevant patterns
to be interpreted, among the many returned, is hard
\cite{chosen,seidl}. We can overcome this problem in three
different ways. First, we can look at the distributions of the MCD
(defined above), the support of the patterns, and the size of the
itemsets to focus our search towards the communities that we consider
relevant, depending on the final application. Figures \ref{fig:supportFreq-dblp}, \ref{fig:quattro}
 report the mentioned distributions (we report the cumulative
 versions, to be able to use the three measures as straightforward
 filters). For better comparison, we reported on the $y$-axes the
 percentage of communities with values of the measures greater than a
 certain thresholds. However, the absolute number of communities can
 be used to choose, depending on the application, the best support,
 size of the itemset, and MCD to select only the relevant
 communities. Second, more generally speaking, the entire
 Constraint-Based Frequent Pattern Mining literature can be applied in
 our scenario at running stage, to drive the search to fewer, more focused, patterns
 \cite{constraints,moreconstraints,bonchiconstraints}. For example, we
 may want only  patterns including or excluding a
 specific dimension, or patterns including dimensions with specific
 properties (e.g., at least 1000 authors). To this extent, it is worth
 noting that MCD is neither (anti-)monotone, nor convertible, nor
 loose-antimonotone. We leave for further
 research the definition of meaningful, application-driven
 constraints, and their effect to the results.
 Lastly, the authors of \cite{chosen} present
 another  methodology for selecting few interesting
 patterns among many, which is not based on constraints. We believe
 that this technique may be also used, and we plan to investigate this
 opportunity in the future.

To  answer \textbf{Q2} we check
whether MCD is correlated to other structural properties
of the nodes of the communities. For example, one possible intuition
is that communities with low density may group together nodes that
were at the borders of the monodimensional communities. To study this, 
 we computed the closeness centrality for each node and for each
 dimension, and checked the correlation between the centrality and the
 density. We did not find any clear sign of direct correlation. 
We checked also for correlation with PageRank, the degree centrality and the betweeness
centrality, for which again we did not have signs of
correlation. Based on these results, we believe that MCD is yet
another measure to be used to filter the results towards more focused results.

% In
%  future studies, we will investigate the relationships with MCD and
%  other measures, or other kind of patterns (e.g., frequent subgraphs).

\begin{figure}[htp!]%
  \begin{tabular}{cc}
\hspace{-4mm}\includegraphics[scale=0.41,clip=true,trim=0 0 10
10]{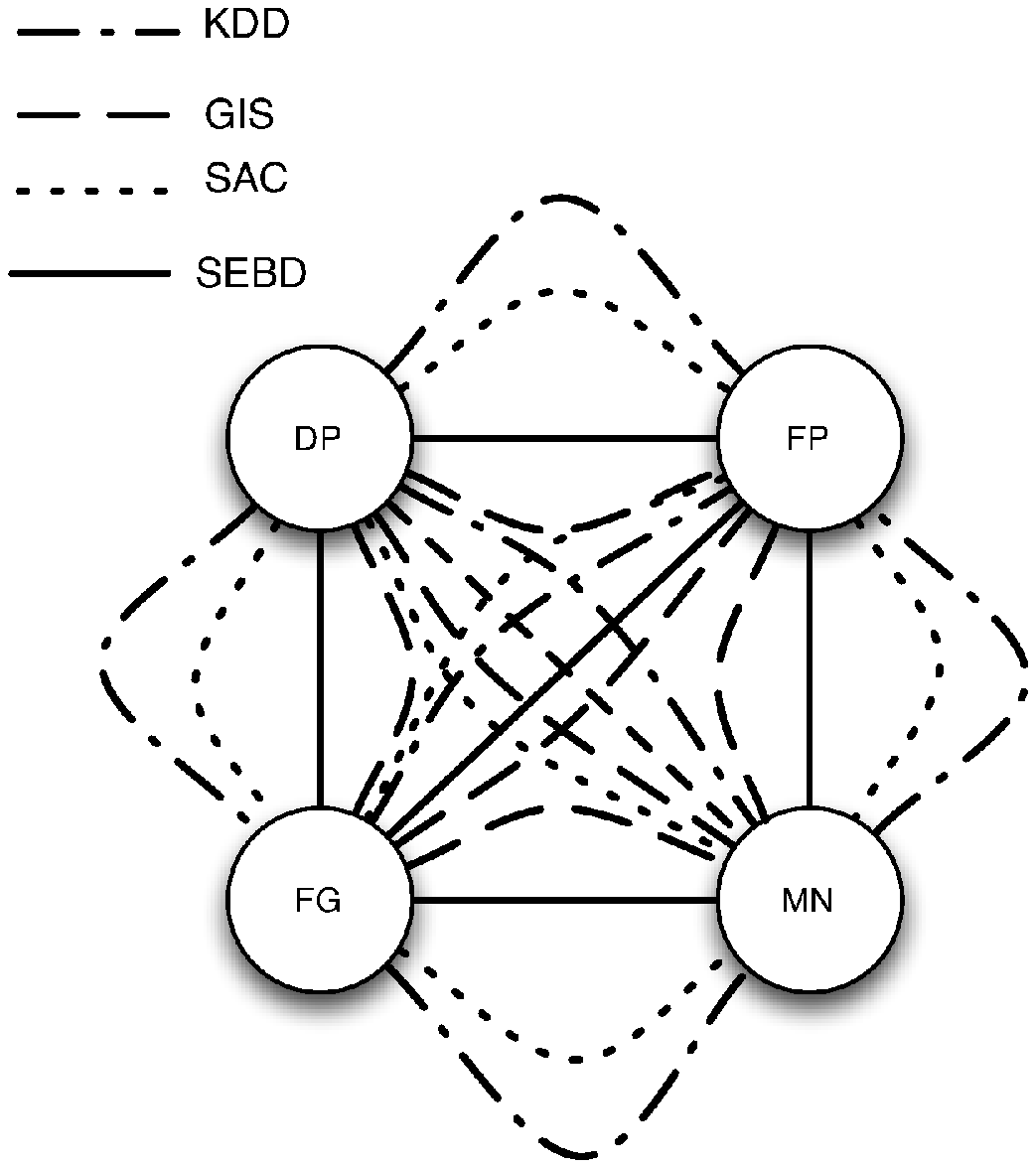} & 
\hspace{2mm}\includegraphics[scale=0.41,clip=true,trim=20 50 10 70]{example-dblp_bw.eps}\\
\hspace{-1mm}(a) MCD=1 in DBLP &
\hspace{2mm}(b) MCD=0.075 in DBLP\\\hline \\
\hspace{-6mm}\includegraphics[scale=0.41]{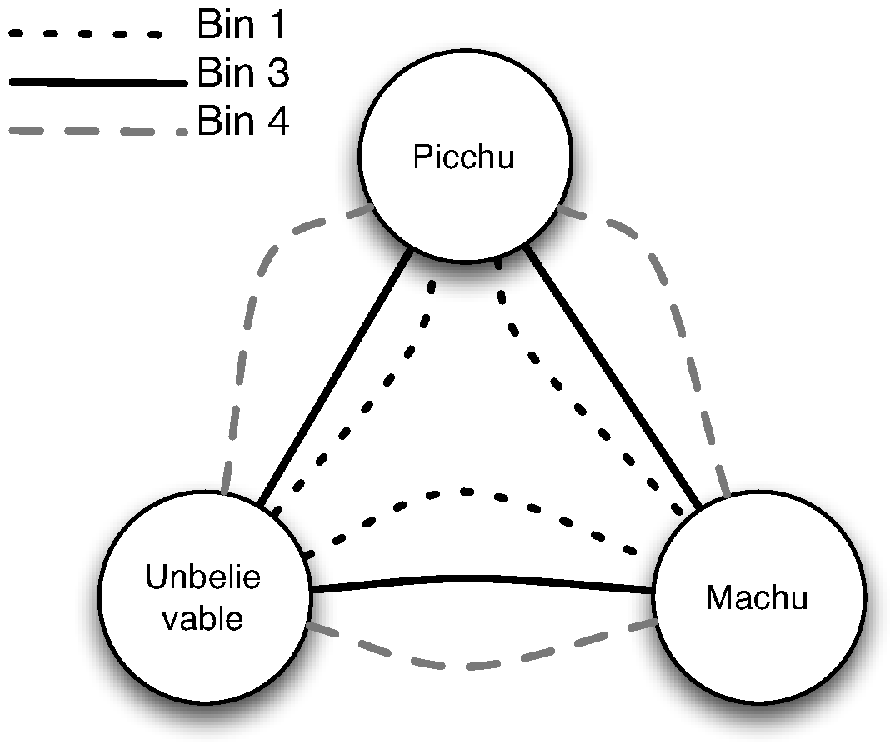} &
\hspace{2mm} \includegraphics[scale=0.41,clip=true,trim=0 0 0 10]{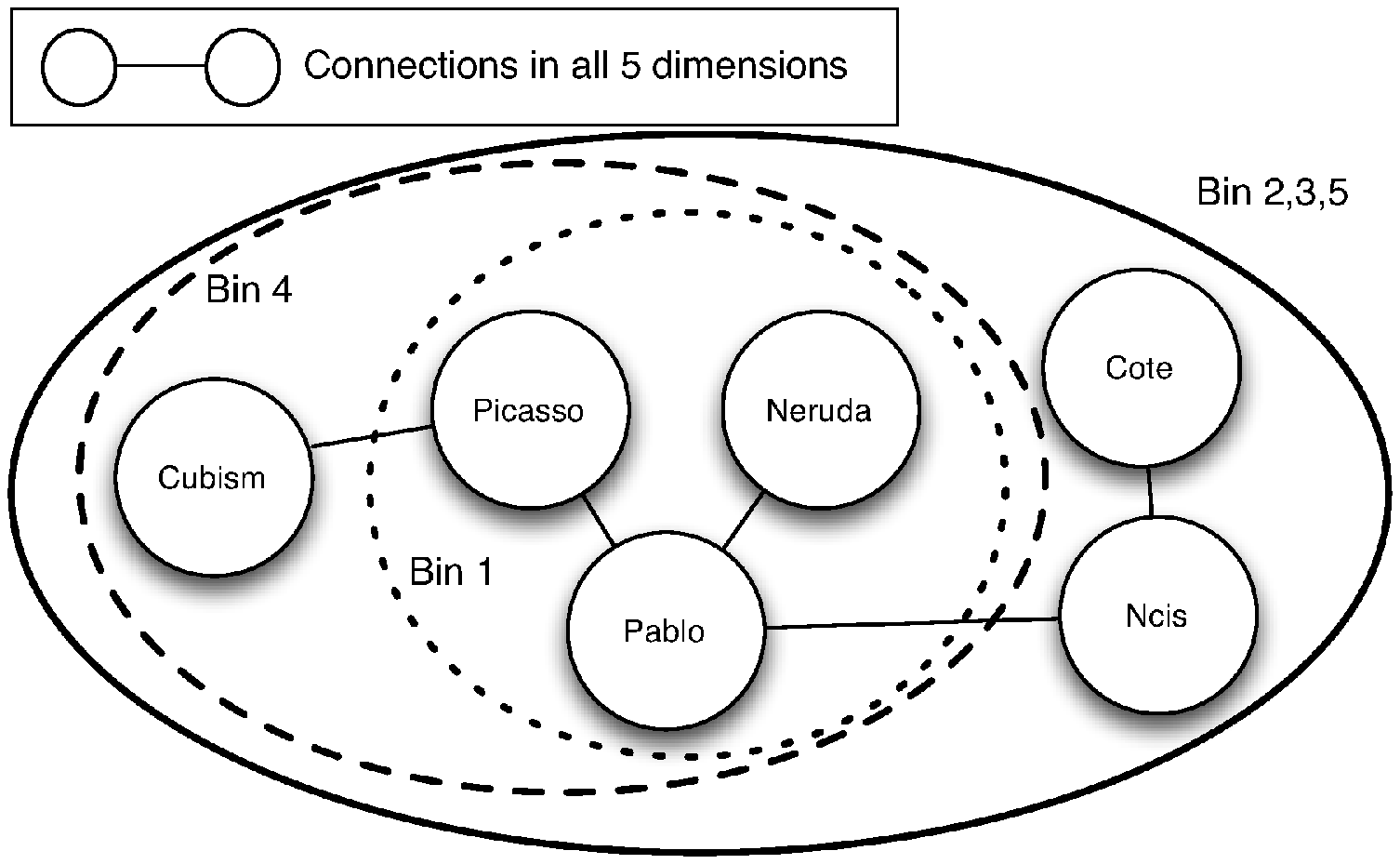} \\

\hspace{-4mm}(c) MCD=1 in Query Log &
\hspace{-1mm}(d) MCD=0.33 in Query Log
  \end{tabular}
\caption{Four communities with high and low MCD extracted from the two
  networks. Nodes in (a): Fosca Giannotti, Mirco Nanni, Dino
  Pedreschi, Fabio Pinelli.  Nodes in (b): Amit Agarwal, Qikai Chen,
  Swaroop Ghosh,  Patrick Ndai,
  Kaushik Roy. The dashed ovals represent the shared
  memberships to the same community in the corresponding dimensions
  (see circle labels). The dashed anonymous nodes in (b) represent
  several nodes belonging to the
  communities in dimensions IOLTS, ISLPED and DATE and are not
  visualized to simplify the readability. In (d), we report with a
  single solid line the point to point connections found in all the
  five dimensions.}
\label{fig:label_tbd}
\end{figure}

Lastly, in order to answer \textbf{Q3}, we extracted a few communities either
minimizing or maximising MCD. In the remainder, we call MCS the number
of nodes in a community. As we have stated above, we can use the
distributions of size, support and MCD to post-process the results to
get only the few interesting ones. We have extracted a few (i.e.,
~200) communities for each network, and we report in Figure
\ref{fig:label_tbd} four of them. Besides the first example, that was
found by searching for one of the co-authors of this paper, the other
ones were found by examining the results filtered by means of the
above mentioned three measures. In particular: Figure
\ref{fig:label_tbd}(b) was found within 260 communities obtained by
constraining $MCD<0.1$, $MCS\geq 2$ and $size\geq 3$; Figure
\ref{fig:label_tbd}(c) was found among 287 communities obtained by
constraining $MCD=1$, $MCS\geq 3$ and $size\geq 2$; Figure
\ref{fig:label_tbd}(d) was found within 286 communities obtained by
constraining $MCD<0.5$, $MCS\geq 4$ and $size\geq 4$. These thresholds
were obtained by looking at the distributions reported above.

Consider the one in Figure
\ref{fig:label_tbd}(a). We discovered a size-4 community connecting FP, FG,
MN and DP with dimensions set $\{KDD, GIS, SAC, SEBD\}$.
It is interesting to observe that, given its very dense connections,
this multidimensional community would have been found also 
by using the methods proposed in
\cite{DBLP:conf/cikm/BerlingerioCG11,onnela}. 

However the method proposed in this paper has the possibility to
discover more complex interactions between dimensions. 
Indeed, the lattice can be used to browse the multidimensional
communities by selecting different dimensions sets. 
To give an example we extracted a size-3 community composed of authors
AA, PN, QC, KR and SG
 with connections in dimensions set $\{IOLTS, DATE, ISLPED\}$, see Figure
\ref{fig:label_tbd}(b) where the different multidimensional
memberships are shown.  
These authors are part of three monodimensional communities, but have
not co-authored papers at these three conferences (there are no links connecting them).
By adding the dimension $ICCD$ (solid line circle), we are able to extract a
size-4 community composed of the first three authors. 
This fourth dimension includes a paper co-authored by the three
authors, which resulted in a $ICCD$-monodimensional community formed
by the three nodes. Interestingly, through this dimension we are able
to specialize the previously discovered 5 authors community. Note that
by using the methods proposed in \cite{DBLP:conf/cikm/BerlingerioCG11,onnela} it
would not be possible to discover how the $ICCD$ dimension could
specialize the community, and so its semantic meaning. This is due to
the fact that more information is included in the results w.r.t. the
mentioned works. 

Similar results can be obtained by applying \ABDUCE to the Query Log
dataset. Finding a community of words in this network means finding a
set of words typically used together in queries that lead to good or
bad results. A set of words found together only in dimension 1 is a set of
words that, used together in a query, lead to very specific results
(users clicked on the first result). Words found together only in dimension
5, on the other hand, lead to lower ranked results. If we find words
together in different dimensions, it may mean that either the concept
the users were looking for is only representable by words used in
conjunction, or that they need more terms to be disambiguated.
An example of the first kind is shown in Figure
\ref{fig:label_tbd}(c), where, maximising the MCD, we were able 
to detect a highly connected multidimensional community where the words \emph{Machu}, 
\emph{Picchu}, and \emph{unbelievable} are connected in three
different dimensions of the dataset. An example of the second kind, on
the other hand, is shown in Figure \ref{fig:label_tbd}(d). In this example, 
we can observe that, if we consider all the dimensions, we obtain a set of words belonging 
to the same multidimensional community with a strong intrinsic semantic correlation 
(i.e. Pablo, Picasso, Neruda -besides sharing their first name, there
exists an edition of a book from Neruda with a Picasso painting on the
cover), removing, then, the most specific dimension  
(Bin 1 -- i.e. click on the first returned result) we include words
that make the concept broader. 
Also in this case, the methods proposed in
\cite{DBLP:conf/cikm/BerlingerioCG11,onnela} do not allow to
investigate  
the effect of the different dimensions on the specialization of the
communities and, thus, the intrinsic semantic correlation among  
different words. 

\subsection{Comparison with previous approaches}\label{sec:comparison}

As reported in Section \ref{sec:related}, in
\cite{DBLP:conf/cikm/BerlingerioCG11}, the authors proposed another 
way to extract multidimensional communities. Their approach is based
however on a different concept of communities: a multidimensional
community groups nodes that are highly multidimensionally
connected. How this multidimensional connectedness is evaluated is
left at the end of the process, by post-processing the resulting
communities. Their approach is composed of the following steps: first,
the multidimensional network is collapsed to a monodimensional one (i.e.,
they follow exactly the opposite of our first step), by weighing the
edges in different ways; second, monodimensional community discovery
is performed on the resulting network; on the resulting communities,
multidimensional connections are restored from the original networks;
the communities are then evaluated by means of multidimensional
measures. 
% \begin{figure}
% \centering
% \includegraphics[scale=0.4]{mono-net_bw.eps} 
% \caption{Run-through example: Monodimensional community discovery
%   applied to a collapsed multidimensional network, where edges are
%   weighted by the number of dimensions.}\label{fig:mononet}
% \end{figure}

% Applying their strategy to our example, we would collapse the
% multidimensional network of our run-through example into a weighted monodimensional as depicted
% in Figure \ref{fig:mononet}.  
% In this example, a CD algorithm will find only one community containing all the
% nodes. Without a manual postprocessing step (e.g., reintroducing all
% the edges, or, equivalently, relabeling the edges to represent the
% original 
% multidimensional information) it would 
%  be impossible to find the subcommunity containing only the nodes
% 1, 3, and 5, which instead is automatically detected using our
% method. 

The approach described in \cite{onnela} works in a similar way,
although it presents some differences. The approach works in two
phases: in the first phase, the adjacency matrices corresponding to
each different dimensions are coupled by connecting, for each entity,
its node representation $i$ in dimension $k'$ to its node
representation $j$ in $k''$. This step is driven by a coupling
parameter $\omega$ which controls the weight of this inter-dimension
connection. This is basically a node-centric
 monodimensional collapsing pre-process on the 
multidimensional information, as opposed to
edge-centric as done in \cite{DBLP:conf/cikm/BerlingerioCG11}. In the
second phase, the authors apply a modularity-driven 
monodimensional community discovery to extract the communities. There
are then two main differences between this baseline and the one
presented in \cite{DBLP:conf/cikm/BerlingerioCG11}: first, the
pre-process step in which the multidimensional information is
collapsed is done at the node level, rather than on the edges; second,
instead of being parametric in the monodimensional community discovery
algorithm, the authors apply a strategy that aims at maximizing a
multidimensional version of the modularity function.

We then compared against both these approaches, using a c++
implementation\footnote{https://code.launchpad.net/louvain} 
of the method presented in \cite{onnela}, and a c++ implementation of
the method presented in \cite{DBLP:conf/cikm/BerlingerioCG11}.

We wanted to compare the three approaches at different
levels. In particular we wanted to answer the following: 
\begin{itemize}
\item[\textbf{Q4.}] Quantitative evaluation: how do the sets of returned
  communities found compare? Can we measure their intersection and the
  number of communities that only our method or a given baseline may find? 
\item[\textbf{Q5.}] Qualitative evaluation: what do the different
  concepts of community look like?
\item[\textbf{Q6.}] Scalability: how do the methods perform on
  networks of different size?
\end{itemize}
In order to address the above, we ran \ABDUCE and the two different
baselines, on several subsets
of the DBLP dataset. Hereafter, we refer to \MCDSOLVER for the method
proposed in \cite{DBLP:conf/cikm/BerlingerioCG11} and to \GL for the
method proposed in \cite{onnela}.

We created two additional (w.r.t. the networks presented in Section
\ref{sec:networks}) sets of networks by
taking incrementally large subsets of DBLP, by taking all the nodes,
edges and dimensions contained in different temporal windows. This
was needed to be able to compare 
against the two different baselines, which present different
scalability in terms of both running times and memory occupation, as
we see in Section \ref{sec:scalability} (in particular, we were not
able to run \GL on large networks). 
The
first set, called ``large nets'' hereafter, consists of 11 networks
corresponding to the single year 2010, the years 2009 and 2010, the years between 2008 and
2010, and so on, up to the years from 2000 to 2010. The second set,
called ``small nets'' hereafter, consists of 11 networks corresponding
to the single year 1990, the years 1989 and 1990, and so on, up to the
years from 1980 to 1990.
\begin{table}[t!]
  \centering\small
  %\begin{tabular}{l|rrr||l|rrr|}
\begin{tabular}{cc}
\rowcolors{2}{gray!40}{white}
\hspace{-3mm} 
    \begin{tabular}{|l|rrr|}
 \rowcolor{gray!10}\hline
Net subset & Dim. & Nodes & Edges\\\hline 
2000-2010 & 2,536 & 558,800 & 2,668,497\\
2001-2010 & 2,477 & 544,608 & 2,585,251\\
2002-2010 & 2,401 & 528,958 & 2,489,462\\
2003-2010 & 2,317 & 511,178 & 2,365,979\\
2004-2010 & 2,244 & 489,228 & 2,216,735\\
2005-2010 & 2,126 & 458,763 & 2,009,903\\
2006-2010 & 1,977 & 423,755 & 1,772,646\\
2007-2010 & 1,848 & 379,182 & 1,496,042\\
2008-2010 & 1,707 & 325,246 & 1,182,161\\
2009-2010 & 1,530 & 260,248 & 840,916\\
2010-2010 & 1,172 & 163,374 & 431,296\\
\hline    \end{tabular}
&
\rowcolors{2}{gray!40}{white}
    \begin{tabular}{|l|rrr|}
 \rowcolor{gray!10}\hline
Net subset & Dim. & Nodes & Edges\\\hline 
1980-1990 & 390 & 37,440 & 71,535\\
1981-1990 & 380 & 36,735 & 69,766\\
1982-1990 & 376 & 35,912 & 67,733\\
1983-1990 & 368 & 34,853 & 65,048\\
1984-1990 & 359 & 33,688 & 61,955\\
1985-1990 & 344 & 31,797 & 57,432\\
1986-1990 & 335 & 29,770 & 52,271\\
1987-1990 & 321 & 26,506 & 44,631\\
1988-1990 & 303 & 23,031 & 36,935\\
1989-1990 & 265 & 18,275 & 27,065\\
1990-1990 & 188 & 11,780 & 15,927\\
\hline
    \end{tabular}
\\\ \\
\multicolumn{1}{c}{(a) Large nets }&
\multicolumn{1}{c}{(a) Small nets }
    \end{tabular}
  \caption{Statistics of the large and small nets used in the comparisons}
  \label{tab:largesmallnet}
\end{table}
 Table \ref{tab:largesmallnet} reports the
basic statistics of the two sets of networks. As we see, the small
nets are much smaller than the large ones, in terms of nodes, edges
and number of dimensions.

\subsubsection{Quantitative evaluation}

\begin{figure*}[th!]
  \centering\small
  \begin{tabular}{cc}
    \hspace{-3mm}    \includegraphics[width=0.51\linewidth]{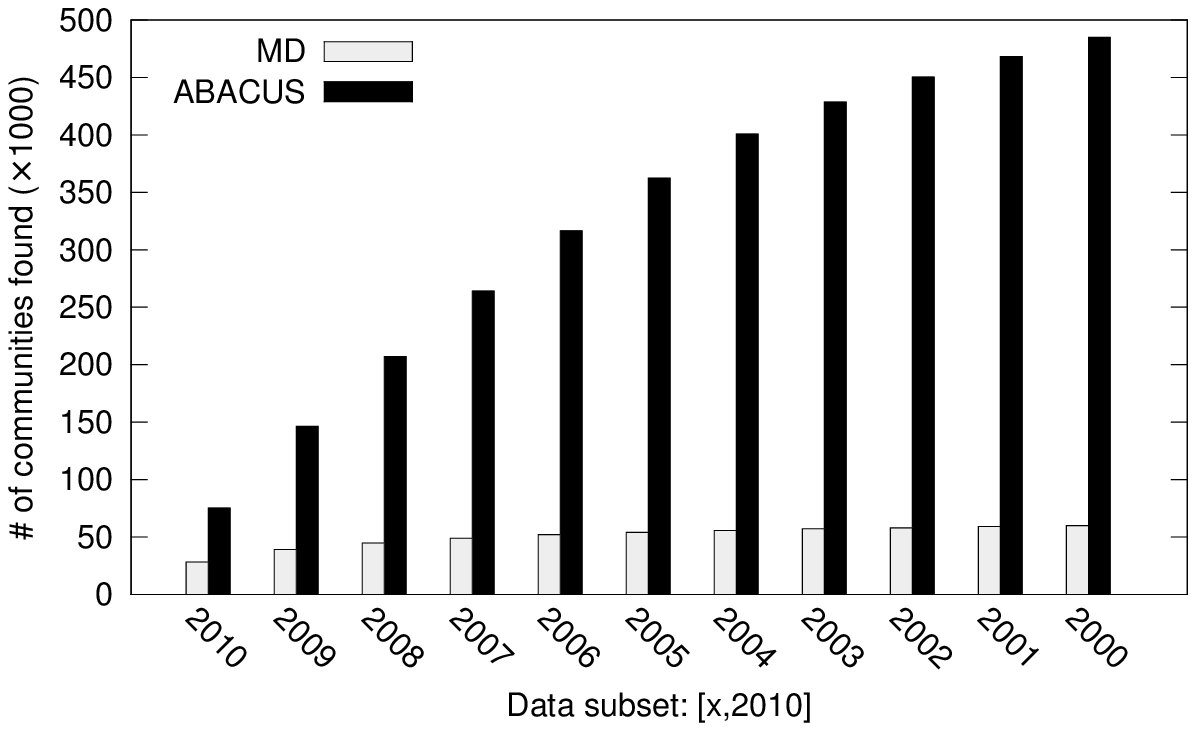} & 
    \hspace{-6mm}    \includegraphics[width=0.51\linewidth]{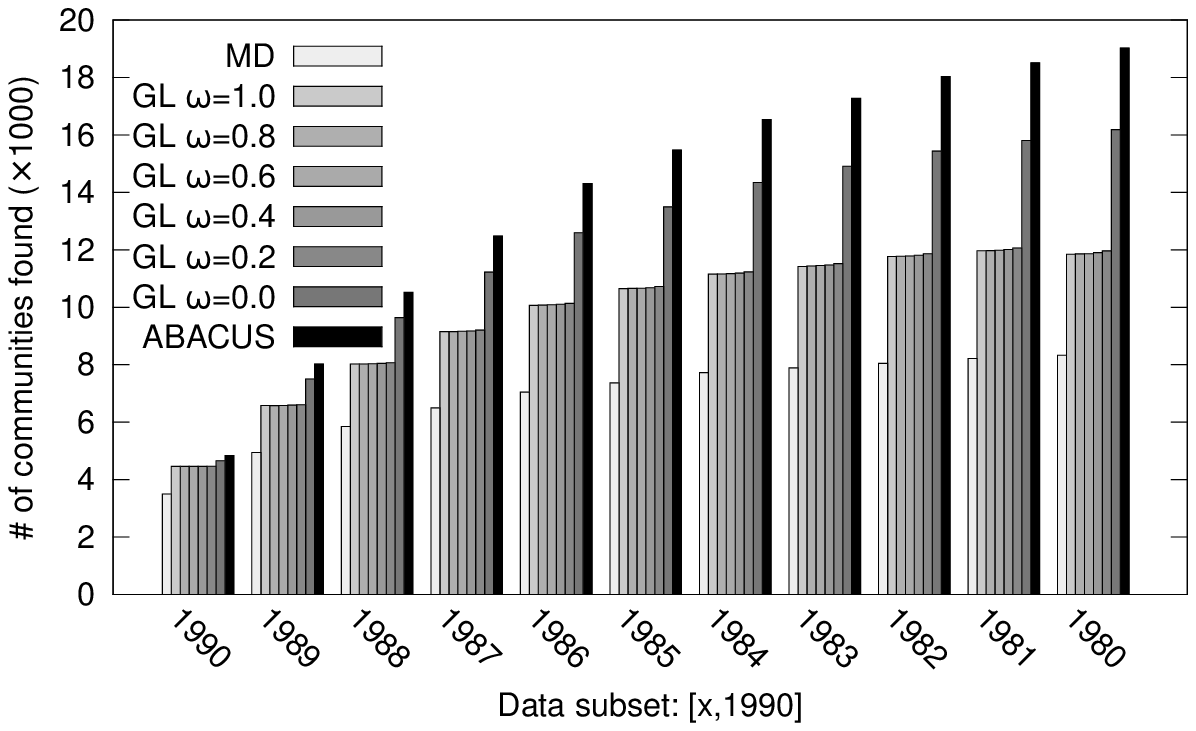} \\
    \hspace{-3mm}    (a) communities found in the large nets &
\hspace{-6mm}     (b) communities found in the small nets \\
  \end{tabular}
  \caption{Quantitative comparisons between \ABDUCE and the baselines 
    (each x value corresponds to an additional year included in the
    subset, from 2000 to 2010 in (a), and from 1980 to 1990 in (b)). 
    Number of communities found by \ABDUCE and \MCDSOLVER on the large
    net in (a), and by \ABDUCE and all the baselines on the small net
    in (b). In these plots, \MCDSOLVER is always the leftmost bar within
  a stack, and \ABDUCE is always the rightmost one.}
  \label{fig:comparisons}
\end{figure*}

Figure \ref{fig:comparisons}(a) reports the number of communities
found  by \ABDUCE and \MCDSOLVER in the large networks, while Figure
\ref{fig:comparisons}(b) reports the number of communities found by
\ABDUCE, \GL with different values of $\omega$ and \MCDSOLVER in the
small networks. 

In the large networks, as we see, due to the strategy of collapsing
the multidimensional network to a monodimensional one, the number of
communities found by \MCDSOLVER becomes nearly stable after adding
four years. In fact, after the first step, each additional year
included into the subset is only changing the weight of existing edges,
instead of creating new ones (and bringing new nodes). On the other
hand, the search space of \ABDUCE grows consistently up to the last two
or three steps, where the growth slows down. By keeping the dimensions
separated, in fact, each additional year is able to provide a
significant number of new combinations to the previous ones. Although
the number of results returned by \ABDUCE is high, we have discussed
in \textbf{Q1} how to deal with it.

In the small networks, the above trend is followed as well, but we
can make further considerations regarding different baselines. First
of all, we see how, due to the definition of the \GL approach, setting
$\omega=0$ leads to a larger number of communities when comparing to
other values of $\omega$. This value of the parameter actually forces
each node representing the same entity in different dimensions to be
grouped separately. In other words, the dimensions are treated in a
disjoint way, i.e. the algorithm performs community discovery in each
dimension separately. 
\begin{figure*}[bh!]
  \centering\small
  \begin{tabular}{c}
    \includegraphics[width=.9\linewidth]{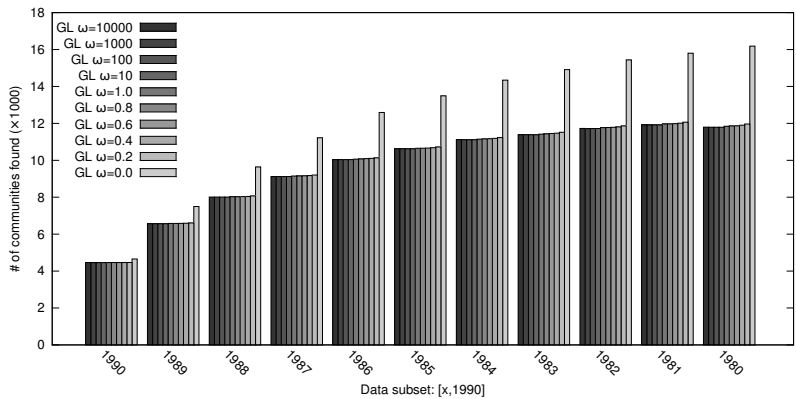} \\
  \end{tabular}
  \caption{Number of communities in the small nets returned by \GL for different values
  of $\omega$.}
  \label{fig:glomega}
\end{figure*}

The plots also shows the difference in
the number of returned communities by values of $\omega$ greater than
zero, although there appear to be no much differences within the
experiments ran with values greater than zero. 
To better explore the sensitivity of these experiments to the $\omega$
parameter we also ran \GL with values up to 10000. Figure
\ref{fig:glomega} shows that there is no substantial difference in
running \GL with values larger than zero and up to 10000. Because of
this, and for sake of simplicity, in the following we only show the
results obtained with a few 
values of $\omega$.

\begin{figure*}[th!]
  \centering
  \begin{tabular}{cc}
    \hspace{-3mm}    \includegraphics[width=0.51\linewidth]{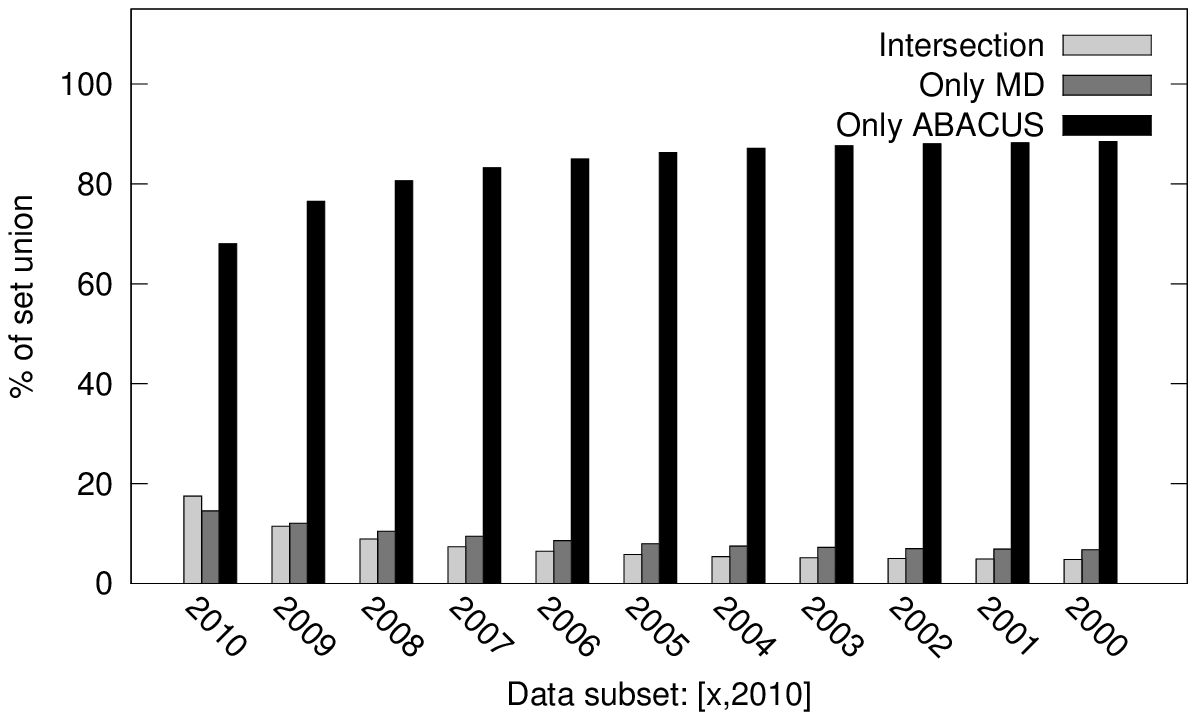} &
    \hspace{-6mm}    \includegraphics[width=0.51\linewidth]{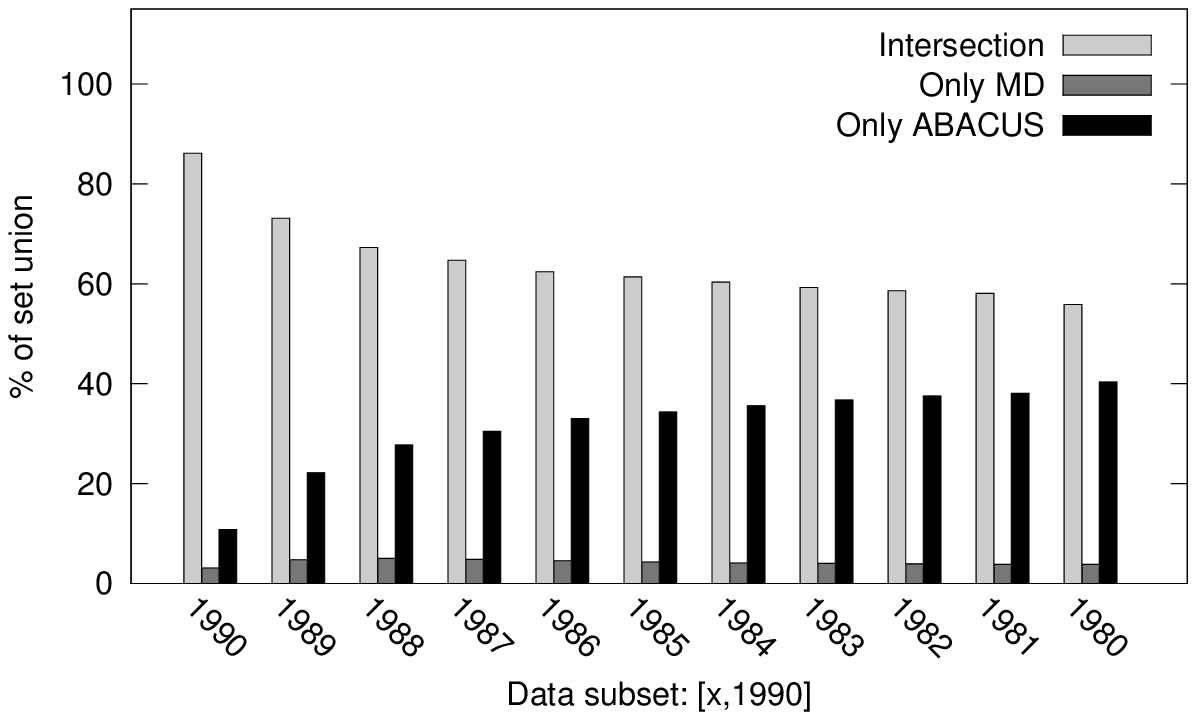} \\
    \hspace{-3mm} (a) large nets, \MCDSOLVER   &
    \hspace{-6mm} (b) small nets, \MCDSOLVER \\
    \hspace{-3mm}    \includegraphics[width=0.51\linewidth]{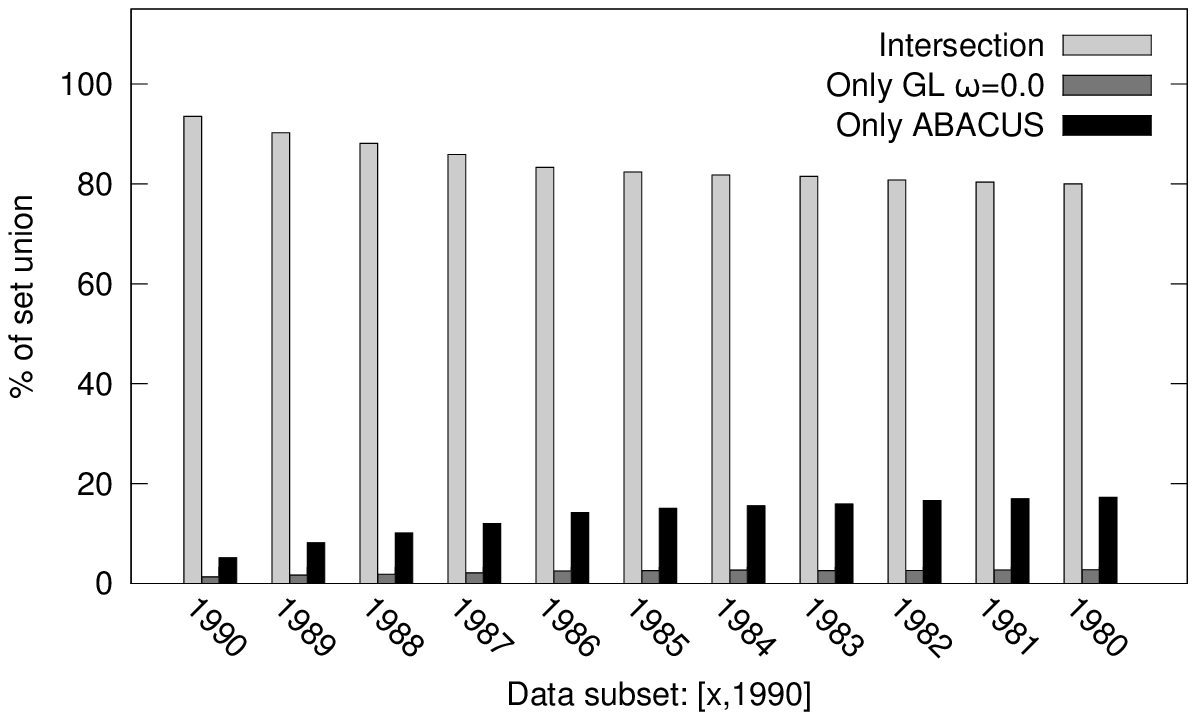} &
    \hspace{-6mm}    \includegraphics[width=0.51\linewidth]{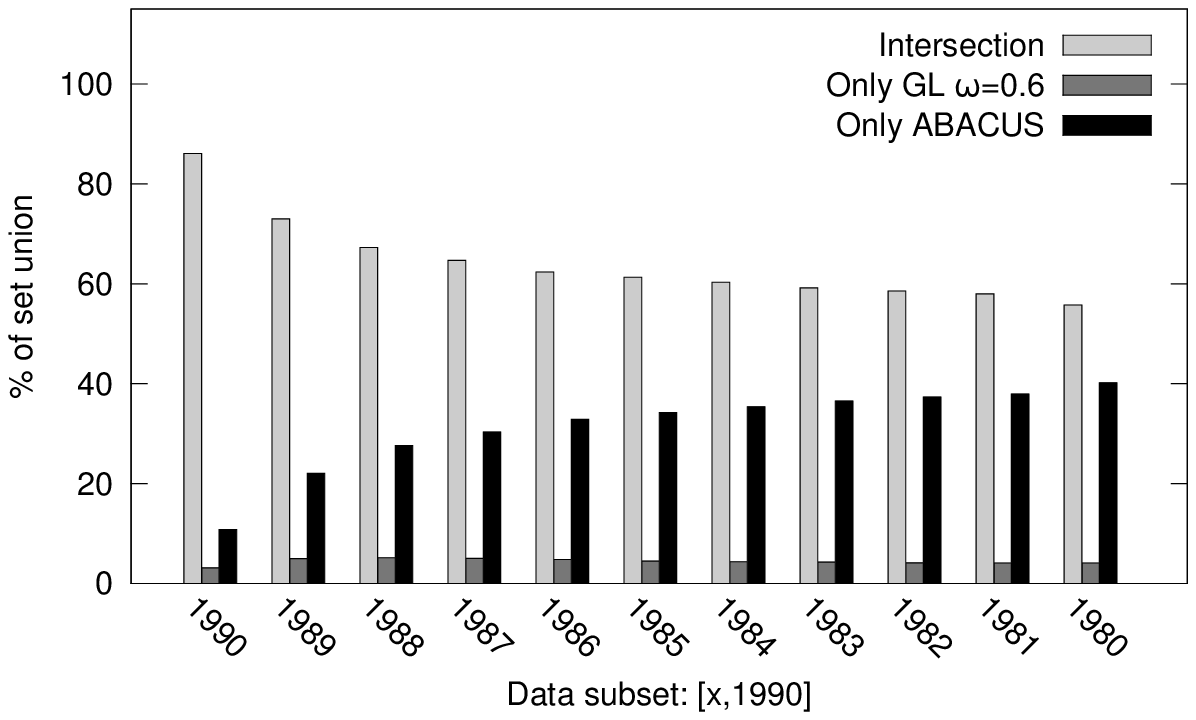} \\
    \hspace{-3mm} (c) small nets, \GL with $\omega=0.0$ &
    \hspace{-6mm} (d) small nets, \GL with $\omega=0.6$  \\
  \end{tabular}
  \caption{Comparison between sets of communities found. Each plot
    compares \ABDUCE against a different baseline or net: large nets and
    \MCDSOLVER in (a), small nets and \MCDSOLVER in (b),
    small nets and \GL with $\omega=0.0$ in (c), small net and
    \GL with $\omega=0.6$ in (d). For each pair of sets of communities A (for
    \ABDUCE) and B (for baseline), we show, for each interval of
    years: $|A\cap B|/|A\cup B|$ -i.e. the portion of
    communities found by both \ABDUCE and the baseline with the light gray
    bar; $|A\setminus B|/|A\cup B|$ -i.e. the portion of communities
    found only by the \ABDUCE- with the black bar; $|B\setminus A|/|A\cup B|$ -i.e. the portion of communities found only by the
    baseline- with the dark gray bar. In these plots, \MCDSOLVER or
    \GL are always the leftmost bar within
  a stack, and \ABDUCE is always the rightmost one.
}
  \label{fig:intersections}
\end{figure*}

% Figure \ref{fig:comparisons}(a) reports the number of communities
% found by the two methods. As we see, due to the strategy of collapsing
% the multidimensional network to a monodimensional one, the number of
% communities found by the baseline becomes nearly stable after adding
% four years. In fact, after the first step, each additional year
% included into the subset is only changing the weight of existing edges,
% instead of creating new ones (and bringing new nodes). On the other
% hand, the search space of \ABDUCE grows consistently up to the last two
% or three steps, where the growth slows down. By keeping the dimensions
% separated, in fact, each additional year is able to provide a
% significant number of new combinations to the previous ones. Although
% the number of results returned by \ABDUCE is high, we have discussed
% in \textbf{Q1} and \textbf{Q1} how to deal with it.

Looking at the plots, a few clear questions arise: are the three methods finding the
same communities? Is one method returning communities found also by
the competitors? Can we identify (classes of) communities that can be
found only by one of the three methods? Figure \ref{fig:intersections}
partially answers these questions from a quantitative point of
view. Calling A the
set of communities found by \ABDUCE and B the set returned a given
baseline, the light gray bar (always the leftmost bar in a stack)
shows $|A\cap B|/|A\cup B|$ -i.e. the portion of
communities found by both, 
the dark gray bar (always the bar in the middle of a stack) shows $|B\setminus A|/|A\cup B|$ -i.e.
the portion of communities found only by the baseline-, and the
black bar (always the rightmost bar in a stack) shows $|A\setminus B|/|A\cup B|$ -i.e. the
portion of communities found only by \ABDUCE. Note that in order to compare the
communities found we had to remove the multidimensional information
contained in those found by \ABDUCE. This step is however correct, i.e. there cannot be two
instances of the same set of nodes tied to two different sets of
dimensions (itemsets) as this would violate the theory behind the
\emph{closed} itemsets. Note also that, in analogy with the majority
of the works on community discovery, and on frequent pattern mining,
we perform \emph{exact} matching here, thus we are only
counting the identical communities in this comparison.

As we see, since the bars report relative
numbers, the ratio of communities that can be found only by the
baselines decreases as the subset of years grows. Put in other words,
even if we know that \ABDUCE is meant to find communities of a
different type than the ones found by the baselines, we see how, for
large datasets, the set of communities found only by one of the baselines
becomes smaller. Moreover, the number of communities that is
found only by \ABDUCE increases accordingly. This is clearly related
to the type of communities that only \ABDUCE can find, i.e. the
communities of unconnected nodes, or, more formally, communities
formed by more than one connected component.
In the following, we answer the above questions
also from a qualitative perspective.

\subsubsection{Qualitative evaluation}
\begin{figure}[t!]
\centering
\begin{tabular}{cc}
\hspace{-8mm}\includegraphics[scale=0.4,clip=true,trim=10 10 30 0]{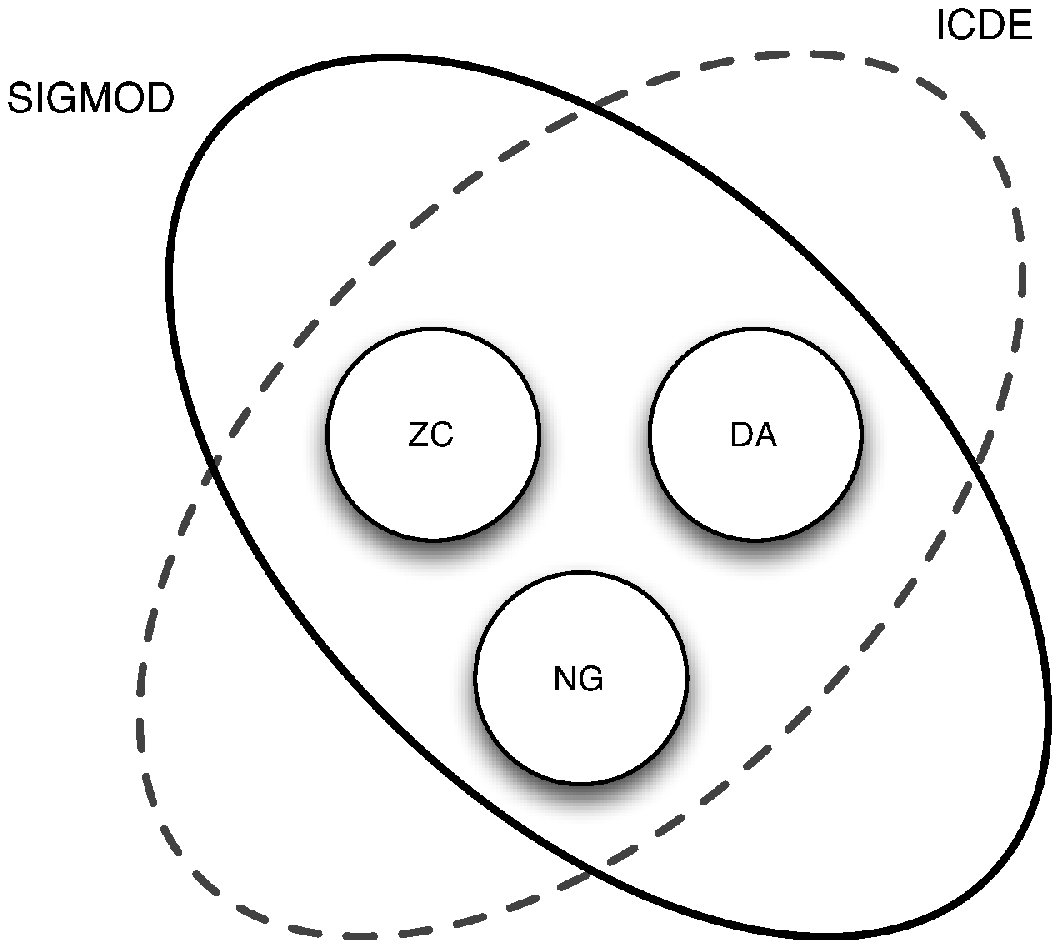}
&
\hspace{3mm}\includegraphics[scale=0.4,clip=true,trim=10 0 00
0]{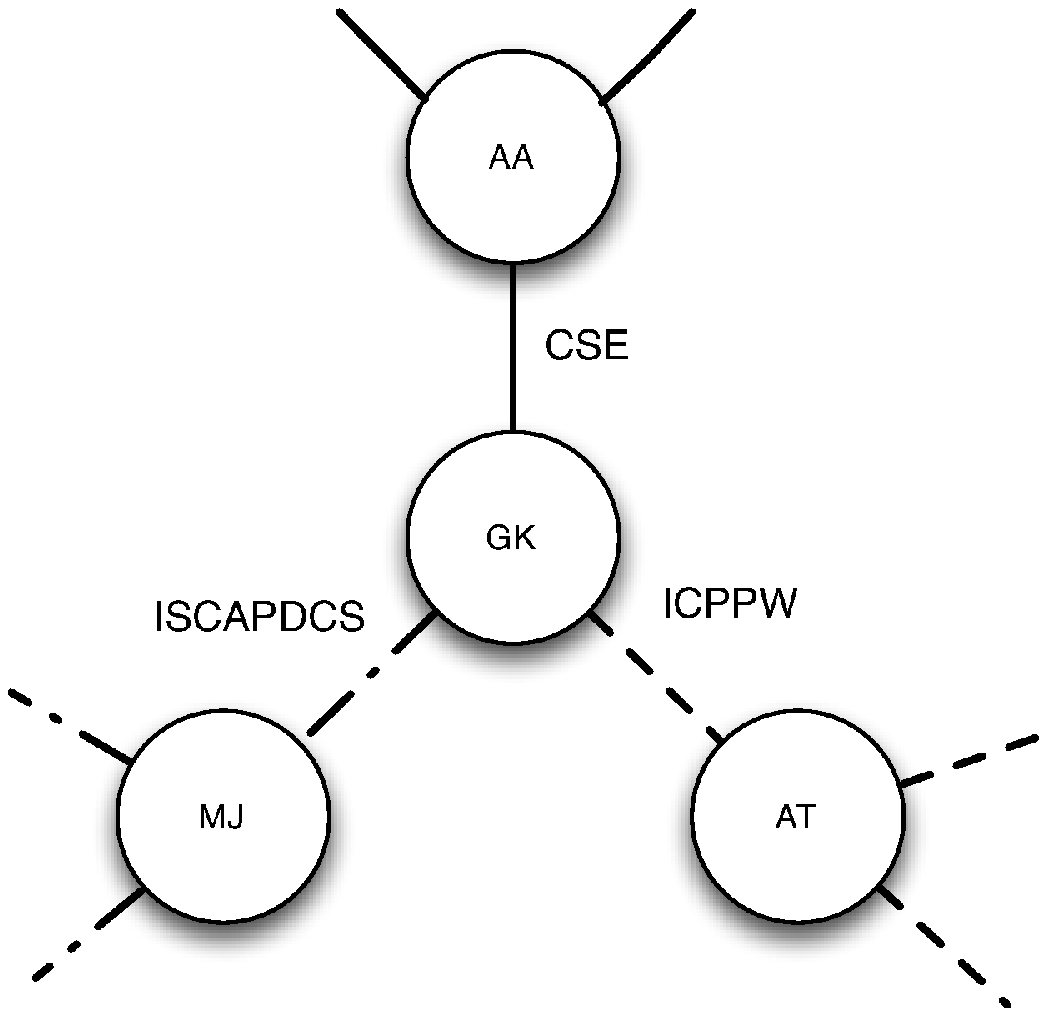}\\
(a) community found & (b)community found \\
 only by \ABDUCE & only by the baselines
\end{tabular}
\caption{Examples of communities found by (a) only \ABDUCE and (b) only
the baselines. Nodes in (a): Deepak Agarwal, Zhiyuan Chen, Nitin
Gupta. Nodes in (b): Anika Awwal, Matthew Jin, Gul N. Khan, Anita TinoIn.
In (b) we also depicted the other outgoing edges from
each nodes that were present in the input data. For the nature of the approaches, (a) could not be
find by the baselines  and (b) could
not be found by \ABDUCE, as different nodes exist in different dimensions.}\label{fig:compex}
\end{figure}
The two concepts of communities found by
\ABDUCE and the baselines are different, without a clear winner (i.e.,
they just reflect different types of interactions among nodes). This
situation can be also detected by the different classes of communities
that only one of the two methods can find. Consider Figure
\ref{fig:dblptoy}: if that was the entire input, \MCDSOLVER would
collapse the network into a monodimensional one and possibly find only
one community containing all the four nodes. \GL would also collapse
the multidimensional connectivity according to the parameter $\omega$.
This cannot happen in
\ABDUCE, as the principle for which the nodes are found in the same
multidimensional community is to share memberships to monodimensional
communities. That is, if Figure \ref{fig:dblptoy}(a) was the entire
input, \ABDUCE would find Jon Doe and John Smith in a multidimensional
community, but not the entire set of nodes, as the remaining two do
not share all the memberships to the other nodes (they do not exist in
dimensions ICDM, CIKM and SIGMOD).
Figure \ref{fig:compex} shows two communities found during our
comparison: (a) was found only by \ABDUCE, and (b) was found only by
both the baselines. Note that we depict all the edges in the original input,
if there were any, and we reported in (b) also the outgoing
edges. While it is clear that (a) cannot be found by the 
baselines (as they rely on connectedness, but there are no edges among
those nodes in the input), in order to confirm that (b) could not be
found by \ABDUCE we had to investigate whether the four nodes were
sharing memberships to the same communities in the depicted
dimensions. That is, even if the image is showing a community that
could not be detected by \ABDUCE if the depicted edges were the entire
input data, there might be in the data other edges (and paths)
connecting the nodes. After post-processing the data, we found that
this was not the case for (b), as different nodes are connected in
different dimensions (see also outgoing edges).

\subsubsection{Scalability}\label{sec:scalability}
% \begin{figure*}[b!]
%   \centering
%   \begin{tabular}{c}
%     \hspace{-4mm}    \includegraphics[width=0.52\linewidth]{runtcomp_bw.eps}\\

%     running time of the two methods applied to the different
%     subsets
%   \end{tabular}
%   \caption{Quantitative comparisons between \ABDUCE and the baseline
%     (each x value corresponds to an additional year included in the
%     subset, from 2010 to 2000): running times of the two
%     approaches, in seconds.}
%   \label{fig:comparisons2}
% \end{figure*}

\begin{figure*}[b!]
  \centering
  \begin{tabular}{cc}
    \hspace{-3mm}    \includegraphics[width=0.51\linewidth]{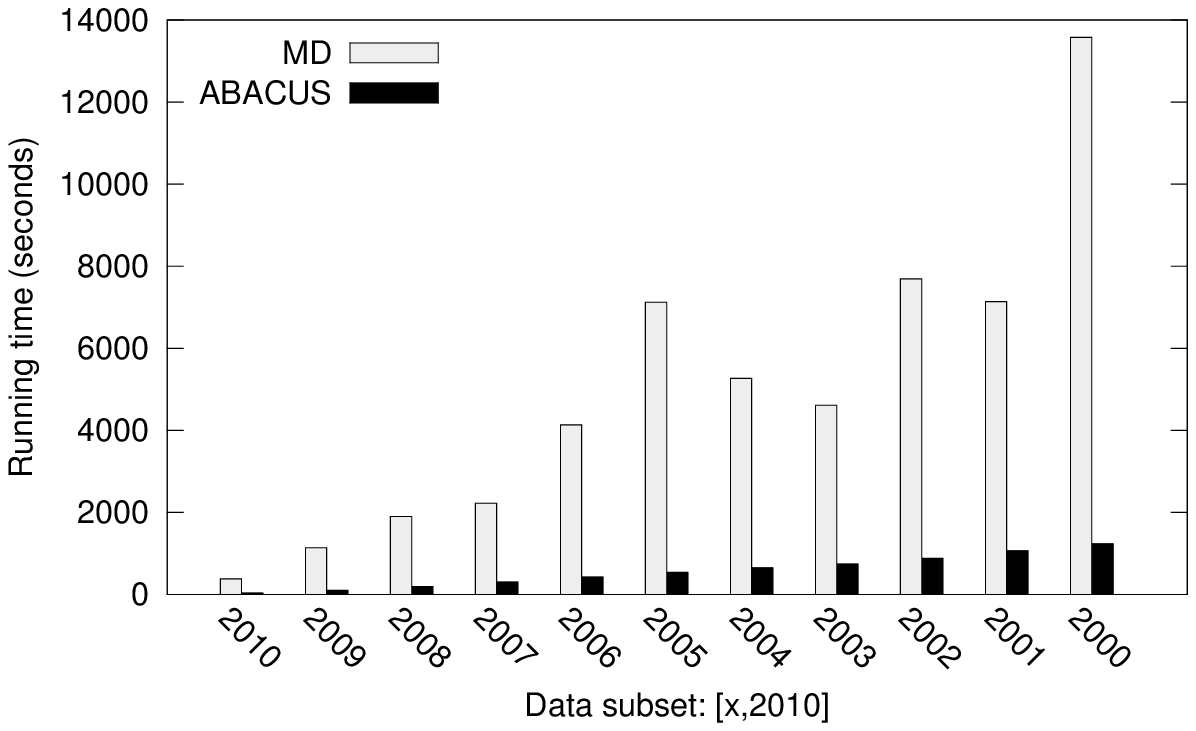} & 
    \hspace{-5mm}    \includegraphics[width=0.51\linewidth]{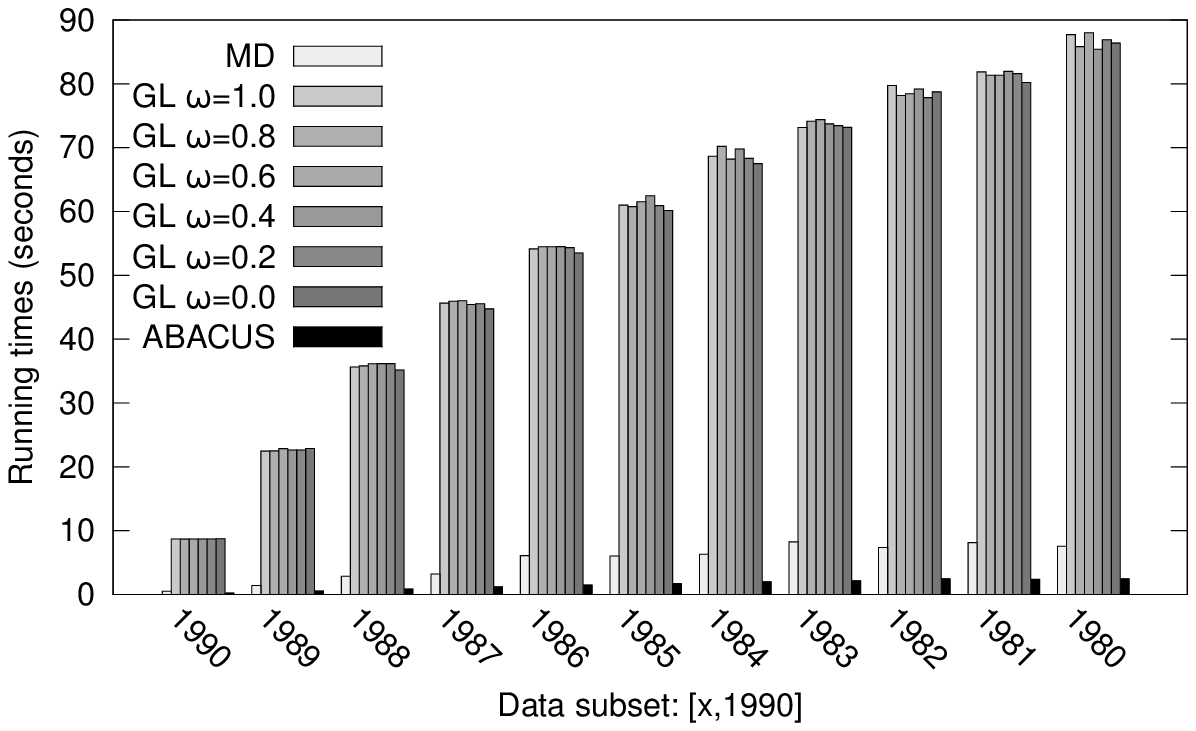} \\
  \end{tabular}
  \caption{Quantitative comparisons between \ABDUCE and the
    baselines. Running times (in seconds) of \ABDUCE and \MCDSOLVER on the large
    net in (a), \ABDUCE and all the baselines on the small nets in (b). }
  \label{fig:runtcomparisons}
\end{figure*}

The last part of our comparison regards scalability. Consider Figure
\ref{fig:runtcomparisons}, reporting the running time (in seconds) of
\ABDUCE and \MCDSOLVER on the large nets on the left, and \ABDUCE,
\MCDSOLVER and \GL on the small nets on the right. As we see, even though by adding years we
implicitly add also dimensions (not all the conferences take place in
all the years, see Table \ref{tab:largesmallnet}), this has a very low impact on the running time of \ABDUCE,
and a very high impact for the baselines.
% This happens despite we pass from roughly 400k edges to 2.7M edges for \ABDUCE,
% and from 370K to 1.9M edges for the baseline (the different number of
% edges for the same subsets is due to collapsing the edges belonging to
% different dimensions).

Note that here we report
 only the running times obtained with a minimum support of two. That
is, we do not test the sensitivity to the 
minimum support parameter, as we already give the worst case. In
reality, if looking for larger 
communities (depending on the application), the running times may be
even lower.

To conclude, \ABDUCE is scalable, and able to process our data in 32 to
1200 seconds (20 minutes) on the large nets, while \MCDSOLVER needed 380 (6 minutes)
to 13500 seconds (225 minutes, i.e. almost 4 hours), and in less than
a second to 2.5 seconds on the small nets, while \MCDSOLVER needed up
to 7.5 seconds and \GL needed up to 86 seconds.

We also report that we were not able to run \GL on networks larger than
the small networks we used because of its memory occupation, that
exceeded 4GB to process the large nets.

\section{Conclusions and future work}\label{sec:conclusions}
In this paper, we have addressed the problem of 
multidimensional community discovery. We have given a definition of multidimensional
community for which nodes sharing memberships to the same monodimensional communities in the different single dimensions are grouped together.
This leads us to define a community extractor combining the use of 
\begin{itemize}
\item
A given monodimensional community discovery algorithm (that could also allow for overlapping communities)
\item
Frequent itemset pattern mining to allow merging discovered monodimensional communities into multidimensional ones
\end{itemize}
By browsing over the lattice generated by the frequent closed itemset
mining algorithm, it is possible to extract multidimensional
communities of different sizes (pattern lengths) and so navigate the
complex multidimensional structure of a network, in a way that
previous methods could not permit. 

The proposed method could lead to the development of analytical tools
to characterize the redundancy in the dimensions, the impact of new
dimensions on the network structure, and more in general to evaluate
the interplay between dimensions. 
For these reasons, we see potential applications in real world problems
including characterizing the interplay between mobility and
communication dimensions in a place-to-place network
\cite{calabrese2011connected}, the similarity between users in a
user-mobility profile network \cite{fabio11}, 
or in the analysis spreading of infectious diseases
\cite{Balcan_Colizza_Goncalves_Hu_Ramasco_Vespignani_2009}. 
We leave for future research the analysis of potential applications of \ABDUCE.

\ \\
\textbf{Acknowledgements.} We would like to thank Mason Porter and
Peter Mucha for helpful discussion around \cite{onnela}, and Vincent
Traag for providing us with a c++ implementation of \GL.

\bibliographystyle{spmpsci}

\end{document}